\documentclass[aps,prc,twocolumn,showpacs,superscriptaddress]{revtex4-2}
\usepackage{natbib}
\usepackage{graphicx}
\usepackage{color}
\usepackage{physics}
\usepackage{float}
\usepackage{natbib}
\usepackage{graphicx}
\usepackage[usenames,dvipsnames]{xcolor}
\usepackage{amsfonts}
\usepackage{amsmath,amssymb}
\usepackage{lineno,soul}
\usepackage{multirow,bigdelim}
\usepackage{cleveref}

\newcommand{\vlwk}{$V_{{\rm low}\mbox{-}k}$}
\newcommand{\thetaeff}{$\Theta_{\rm eff}$}
\newcommand{\thetaeffs}{$\Theta_{\rm eff}$s}

\newcommand{\zbb}{$0\nu\beta\beta$}

\newcommand{\heff}{$H_{\rm eff}$}
\newcommand{\heffs}{$H_{\rm eff}$s}
\newcommand{\qbox}{$\hat{Q}$~box}
\newcommand{\tbox}{$\hat{\Theta}$~box}
\newcommand{\nme}{$M^{0\nu}$}
\newcommand{\nmes}{$M^{0\nu}$s}

\begin{document}

\title{Realistic shell model for ordinary muon capture of $sd$-shell nuclei}

\author{S. L. Lyu}
\affiliation{Dipartimento di Matematica e Fisica, Universit\`a degli
  Studi della Campania ``Luigi Vanvitelli'', viale Abramo Lincoln 5 -
  I-81100 Caserta, Italy}
\affiliation{Istituto Nazionale di Fisica Nucleare, \\ 
Complesso Universitario di Monte  S. Angelo, Via Cintia - I-80126 Napoli, Italy}
\author{G. De Gregorio}
\affiliation{Dipartimento di Matematica e Fisica, Universit\`a degli
  Studi della Campania ``Luigi Vanvitelli'', viale Abramo Lincoln 5 -
  I-81100 Caserta, Italy}
\affiliation{Istituto Nazionale di Fisica Nucleare, \\ 
Complesso Universitario di Monte  S. Angelo, Via Cintia - I-80126 Napoli, Italy}
\author{T. Fukui}
\affiliation{Faculty of Arts and Science, Kyushu University, Fukuoka, 819-0395, Japan}
\affiliation{RIKEN Nishina Center, Wako, 351-0198, Japan}
\author{N. Itaco}
\affiliation{Dipartimento di Matematica e Fisica, Universit\`a degli
  Studi della Campania ``Luigi Vanvitelli'', viale Abramo Lincoln 5 -
  I-81100 Caserta, Italy}
\affiliation{Istituto Nazionale di Fisica Nucleare, \\ 
Complesso Universitario di Monte  S. Angelo, Via Cintia - I-80126 Napoli, Italy}
\author{L. Coraggio}
\affiliation{Dipartimento di Matematica e Fisica, Universit\`a degli
  Studi della Campania ``Luigi Vanvitelli'', viale Abramo Lincoln 5 -
  I-81100 Caserta, Italy}
\affiliation{Istituto Nazionale di Fisica Nucleare, \\ 
Complesso Universitario di Monte  S. Angelo, Via Cintia - I-80126 Napoli, Italy}

\begin{abstract}
We report about a study of the ordinary muon capture in nuclei belonging to the $sd$ shell, an electroweak process that occurs with exchange momenta far larger than ordinary $\beta$ decays ($\approx$ 100 MeV).
Such a characteristic places this transition in an energy range that is consistent with the neutrinoless double-$\beta$ decay, and represents an interesting test for nuclear models to support their predictions of the nuclear matrix elements for such an unobserved process.
For the first time, the calculations are carried out within the realistic shell model (RSM), namely employing effective shell-model Hamiltonians and decay operators derived from realistic nuclear forces, without resorting to any empirical adjustment of the coupling constants.
This is a chapter of a research program that is aimed to assess the realistic shell model in reproducing the observables related to electroweak processes in nuclei, and then to evaluate the reliability of nuclear matrix elements for the neutrinoless double-$\beta$ decay that are calculated within this approach.
We calculate the partial capture rates for many nuclear systems in the $sd$-shell region, as well as their spectroscopic properties, and compare the results with the available experimental counterparts.
Such a comparison tests the relevance of a microscopic approach to the renormalization of transition operators to reproduce data and provide solid predictions of unknown observables.
\end{abstract}

\maketitle

\section{Introduction}
\label{intro}
The study of electroweak processes occurring in nuclear systems is currently a topic of paramount interest for the nuclear structure community, since they are a relevant testing ground for the validation of calculated nuclear matrix elements for the neutrinoless double-$\beta$ (\zbb) decay.
An important issue that has to be met in the calculation of the nuclear matrix elements of electroweak processes is the renormalization of the transition operators, an operation that is necessary to account for the constraints on the degrees of freedom that can be employed to construct nuclear wave functions in a non-relativistic low-energy depiction of atomic nuclei.
This is usually tagged as the so-called {\it quenching puzzle}, namely the need to employ an empirical value of the axial coupling constant $g_A$, that has to be reduced by a factor $q$ smaller than unity \cite{Suhonen13,Ejiri83,Martinez-Pinedo96,Barea15,Suhonen19}.

Previous studies \cite{Arima73,Arima74,Towner87,Arima88} have found that there are two main sources for this overestimation of the Gamow-Teller (GT) transition rates, and they are rooted in the limits of the depiction of the nucleus as a non-relativistic system of interacting nucleons in a low-energy regime.

First, nucleons are not elementary particles, and hence one has to consider effectively their quark structure through the effects of meson-exchange currents (two-body and many-body electroweak currents) \cite{Arima73,Towner83,Baroni16a}.
Second, only few-body methods and {\it ab initio} approaches for nuclear structure calculations start from the solution of the nuclear many-body problem in the full Hilbert space of the configurations. 
Most of nuclear structure models consider a reduced model space, with a restricted number of the degrees of freedom, where the nuclear wave functions are constructed.

To compensate for this, one often resorts to a renormalization of the nuclear Hamiltonian, as well as of any decay operator $\Theta$, and the latter operation translates in the introduction of empirical effective charges for protons and neutrons -- for the nuclear matrix elements of electric transitions between nuclear states --, or of empirical quenching factors for spin- and spin-isospin dependent transitions – namely for magnetic or GT decays.

Our approach to the renormalization of nuclear operators is to derive effective ones for shell-model (SM) calculations, starting from realistic nuclear forces and bare decay operators and employing the many-body perturbation theory to obtain the effective Hamiltonian \heffs~ as well as decay operators \thetaeffs~ \cite{Coraggio20c}.

During last years, we have carried out an extensive study of electroweak decays in nuclear systems, namely the calculation of the nuclear matrix elements of single- and double-$\beta$ decays \cite{Coraggio17a,Coraggio19a,Coraggio22a,Coraggio24a}, as well as the investigation of the sensitivity to effective decay-operators of the energy spectra of forbidden $\beta$ decays \cite{DeGregorio24}.
The goal of such a research activity is to test our theoretical framework, where the calculation of nuclear matrix elements of electroweak decays is performed within a fully microscopic approach, namely without resorting to empirically-adjusted parameters, and the results depend only from the input nuclear potential.
This aims to validate our approach with respect to the calculation of trustable nuclear matrix elements \nme~ for the \zbb~ decay, a second-order electroweak process which has not yet been observed and that is the rarest decay within the Standard Model.

In this work, we extend our investigation of nuclear electroweak processes to the study of ordinary muon capture (OMC)~\cite{Measday01}, which involves a release of energy that is much larger than ordinary $\beta$-decay transitions, since the muon mass is more than two orders of magnitude larger than the electron one.
In particular, the transfer momentum -- that is about 100 MeV/c and mainly absorbed by the emitted neutrino \cite{Measday01} -- is consistent with the one involved in the \zbb~ decay, then the description of OMC by nuclear theoretical models is a challenge that should be accepted to test the reliability of calculated \nmes.

This connection between OMC and \zbb~ decay has ignited a renewed interest in the communities both of nuclear structure theory \cite{Kortelainen02,Kortelainen04,Jokiniemi19,Jokiniemi21,Jokiniemi24,Civitarese25,Simkovic20} and experimental neutrino physics \cite{Gorringe94,Zinatulina18,Hashim23,Araujo24}, and provides another brick in the structure which holds up the research towards the understanding of the nature of the neutrino, as well as of the boundaries of the Standard Model.

In this work, we carry out for the first time a calculation of the nuclear matrix elements of OMC through the realistic shell model (RSM), namely, starting from realistic nuclear potentials, we derive effective SM decay operators and Hamiltonians for nuclei in the $sd$-shell region by way of the many-body perturbation theory
\cite{Kuo81,Suzuki95,Coraggio12a,Coraggio20c}. 
We employ two different realistic nuclear Hamiltonians, since we want also to investigate about the correlation between the quality of the reproduction of the observed spectroscopy and the comparison between calculated and experimental observables which are related to OMC matrix elements.

The two Hamiltonians we have considered are:
\begin{itemize}
\item A low-momentum two-nucleon ($2N$) potential derived from the CD-Bonn high-precision potential \cite{Machleidt01b}, whose repulsive high-momentum components have been renormalized using the \vlwk~ procedure \cite{Bogner02}.
\item A nuclear Hamiltonian based on chiral perturbation theory (ChPT) \cite{Epelbaum09,Machleidt11}, that consists of a high-precision $2N$ potential derived at next-to-next-to-next-to-leading order (N$^3$LO) \cite{Entem02}, and a three-nucleon ($3N$) component at N$^2$LO in ChPT \cite{Navratil07a}.
\end{itemize}

We have found that the results of our calculated spectroscopic properties and OMC nuclear matrix elements depend noticeably upon the input realistic Hamiltonian, which seems to correlate the reproduction of the observed shell evolution in the $sd$-shell region, and the comparison with experimental OMC partial capture rates.

This paper is organized as follows.
\noindent
In Sec. \ref{outline} we outline the derivation of the effective SM Hamiltonian and decay operators, as well as the formalism of the calculation of the single-particle matrix elements of the OMC operator.

The results of the SM calculations are discussed and compared with the available experimental data in Sec. \ref{results}. 
First, we check our nuclear wave functions by comparing the calculated low-energy spectra, electromagnetic transition strengths and moments, and Gamow-Teller (GT) log$ft$s of parent and daughter nuclei, which are involved in the OMC under consideration, with their experimental counterparts.
In particular, we have also considered those features that could be considered as the signatures of the shell evolution of $sd$-shell nuclei, namely the oxygen two-neutron separation energies $S_{2n}$ – which characterize the dripline of these isotopes --, the behavior of the yrast $J=2^{+}$ excitation energies for the same class of nuclear systems, and the evolution of the effective single-particle energies (ESPE), that reflects the properties of the monopole components of the SM Hamiltonian \cite{Utsuno99,Otsuka22}.
Then, we report our calculated OMC partial capture rates and compare them with available data.
As already mentioned, we focus our attention on the correlation between the quality of the reproduction of spectroscopic properties and the one of partial capture rates, which are sensitive to the chosen \heff. 

In the last section (Sec. \ref{conclusions}), we summarize the conclusions of this study in connection with the assessment of the realistic shell model to provide reliable nuclear matrix element, and present some perspectives of our future efforts in the investigation about the OMC process.

\section{Outline of the theory}\label{outline}

\subsection{The effective SM Hamiltonian}\label{effham}
In this section we are going to sketch out briefly our approach to the derivation of the effective SM Hamiltonian, a procedure that we have reported in two recent works \cite{Coraggio24a,DeGregorio24}.
More details can be found in the topic paper in Ref. \cite{Coraggio20c}.

As mentioned in the Introduction, we construct two \heffs.

One \heff~ is derived from the high-precision CD-Bonn $2N$ potential
\cite{Machleidt01b}, then the non-perturbative repulsive high-momentum
components are integrated out, by way of the \vlwk~ unitary transformation \cite{Bogner02,Coraggio09a}. 
This renormalization procedure provides a smooth $2N$ potential which preserves all the two-nucleon observables as provided by the CD-Bonn one.
As in our recent papers \cite{Coraggio16a,Coraggio17a,Coraggio19a,Coraggio20a,Coraggio22a,DeGregorio24}, we have chosen a ``hard’’ value of the cutoff, being $\Lambda=2.6$ fm$^{-1}$, since we have found that the larger the cutoff the smaller the role of the missing three-nucleon force \cite{Coraggio15b}.

The other \heff~ is obtained from the high-precision $2N$ potential developed by Entem and Machleidt through a chiral perturbative expansion at
N$^3$LO \cite{Entem03}, introducing a regulator function whose cutoff parameter is  $\Lambda = 500$ MeV, and that is characterized by a smooth behavior in the high-momentum regime and can be profitably employed for a perturbative derivation of \heff~ (see the contents of Sec. II in Ref. \cite{Ma19}).
It should be pointed out that a main advantage of the effective field theory (EFT) is to perform a  derivation of a nuclear Hamiltonian by introducing two- and many-body forces on an equal footing \cite{Weinberg92,vanKolck94,Machleidt11}, since most
interaction vertices that appear in the three-nucleon force (3NF) and
in the four-nucleon force (4NF) also occur in the two-nucleon one
(2NF).
Then, aside a $2N$ component – derived at N$^3$LO in ChPT – we include a $3N$ term, which is derived at N$^2$LO in the chiral perturbative expansion.
The latter comprises three 3NF topologies: the two-pion exchange (2PE), one-pion exchange (1PE), and three-nucleon-contact interactions \cite{Machleidt11}, which are characterized by a set of low-energy constants (LECs) that appear already in the 2PE component of the 2NF.
The 3NF 1PE contribution contains a new LEC $c_D$, and another one -- $c_E$ -- characterizes the 3NF contact potential.
Since $c_D$, $c_E$ LECs do not appear in the $2N$ problem, the renormalization procedure requires that they should be fixed to reproduce the observables of $A\geq 3$ systems.

In present work, we have adopted the $c_D, c_E$ values as suggested in the work by Navratil {\it et al} in Ref. \cite{Navratil07a}, where the authors first constrained the relation of  $c_D$-$c_E$, and then investigated a set of observables in light $p$-shell nuclei to give a second constraint.
This parametrization, namely $c_D=-1$ and $c_E=-0.34$, has been employed also in our preceding works, where this nuclear Hamiltonian has been considered \cite{Fukui18,Ma19,Coraggio20e,Coraggio21, Coraggio24a}.

It is worth pointing out that the N$^2$LO $3N$ component plays a crucial role to reproduce the shell evolution and closures, as pointed out in Refs. \cite{Ma19,Fukui24,Coraggio24b}

Finally, the Coulomb potential is explicitly included in the proton-proton channel of the nuclear Hamiltonian.

The $2N$ and $3N$ matrix elements are chosen as the interaction vertices of a perturbative expansion of \heff, and an extended description of the many-body perturbation theory approach to the nuclear \heff~ can be found in Refs. \cite{Hjorth95,Coraggio12a,Coraggio20c}, so here we only highlight the procedure that we have followed.

The starting point is to consider the nuclear Hamiltonian $H$ for $A$ interacting nucleons, which, according to the shell-model ansatz, is split into a one-body term $H_0$, whose eigenvectors set up the SM basis, and a
residual interaction $H_1$, through the introduction of an auxiliary potential $U$:

\begin{eqnarray}
 H &= & T + V_{NN} = (T+U)+(V_{NN}-U) \nonumber\\
~& = &H_{0}+H_{1}~,\label{smham}
\end{eqnarray}

\noindent
where the auxiliary potential $U$ is chosen to be the harmonic-oscillator (HO) one, with a value of the HO parameter $\hbar \omega = 14$ MeV.

Obviously, the eigenvalue problem of $H$ for a many-body system, and
in an infinite Hilbert-space of $H_0$ eigenvectors, cannot be solved, leading to the necessity to derive an effective Hamiltonian through a similarity transformation \cite{Suzuki80,Stroberg19}, that projects the eigenvalue problem to a truncated model space, in this case spanned by three proton and neutron orbitals $0d_{5/2}, 0d_{3/2}, 1s_{1/2} $, outside $^{16}$O doubly-closed core.

We derive \heff~ by way of the time-dependent perturbation theory, performing the Kuo-Lee-Ratcliff folded-diagram expansion in terms of the $\hat{Q}$-box vertex function \cite{Kuo90,Hjorth95,Coraggio12a}:

\begin{equation}
H^{\rm eff}_1 (\omega) = \hat{Q}(\epsilon_0) - P H_1 Q \frac{1}{\epsilon_0
  - Q H Q} \omega H^{\rm eff}_1 (\omega) ~,\label{eqfinal}
\end{equation}
\noindent
where $\omega$ is the wave operator decoupling the model space $P$ and
its complement $Q$, and $\epsilon_0$ is the eigenvalue of the unperturbed
degenerate HO Hamiltonian $H_0$.

The \qbox~is defined as
\begin{equation}
\hat{Q} (\epsilon) = P H_1 P + P H_1 Q \frac{1}{\epsilon - Q H Q} Q
H_1 P ~, \label{qbox}
\end{equation}
\noindent
and $\epsilon$ is an energy parameter called the ``starting energy''.

The exact calculation of the \qbox~ cannot be carried out, then the
term $1/(\epsilon - Q H Q)$ is expanded as a power series

\begin{equation}
\frac{1}{\epsilon - Q H Q} = \sum_{n=0}^{\infty} \frac{1}{\epsilon -Q
  H_0 Q} \left( \frac{Q H_1 Q}{\epsilon -Q H_0 Q} \right)^{n} ~.
\end{equation}

Then, we perform an expansion of the \qbox~up to the third order in perturbation theory (n=1) \cite{Coraggio20c}, and the set of two-body configurations belonging to the subspace $Q$ is characterized by an excitation energy smaller than $E_{\rm max}= N_{\rm max} \hbar \omega$, where $N_{\rm max} = 20$ \cite{Coraggio12a}.

This  is sufficient to obtain convergent values of the single-particle (SP) energies and two-body matrix elements of the residual interaction (TBMEs), as it has been shown in Refs. \cite{Coraggio18,Ma19}.

The perturbative calculation of the \qbox~ allows then solving the non-linear matrix equation (\ref{eqfinal}) and obtain \heff~ by way of iterative techniques \cite{Krenciglowa74,Suzuki80}, or graphical non-iterative methods \cite{Suzuki11}.

Since the nuclei that are under investigation are characterized by a large
number of valence nucleons, we have included contributions from
induced three-body forces in the calculation of the \qbox, that
involve also three valence nucleons, resorting to a normal-ordering decomposition of the $3N$ induced-force contributions arising at second order in perturbation theory.
Then, we retain only the two-body term that is density-dependent from the
number of valence nucleons, and more details can be found in Refs. \cite{Coraggio20c,Coraggio20e}.
The reference state in such a procedure is chosen to be the ground state of the nucleus under study, and a fractional filling of model-space orbitals is considered, as also suggested in Ref. \cite{Stroberg17} for the application of the valence-space in-medium similarity transformation group (VS-IMSRG) approach.

The normal-ordering procedure has been adopted also to calculate the  contributions at first order in many-body perturbation theory for the calculation of the \qbox~ of the N$^2$LO $3N$ component of the ChPT Hamiltonian \cite{Fukui18}

The SM parameters, namely the SP energies and the TBMEs of the
residual interaction, are reported -- for both \vlwk~ and ChPT \heffs~ -- in the Supplemental Material \cite{supplemental2025}.

\subsection{Effective shell-model decay operators}\label{effopsec}
The necessity to employ effective operators is rooted in the issue that the diagonalization of the \heff~ does not provide the true nuclear wave-functions, but their projections onto the model space $P$.
Then, as it happens for the SM Hamiltonian, any decay operator $\Theta$ has to be renormalized, so that the effective SM operator \thetaeff~ accounts for the neglected degrees of freedom belonging to the $Q=1-P$ subspace.

In this work, as in other previous study of nuclear electroweak transitions \cite{Coraggio17a,Coraggio19a,Coraggio20a,Coraggio22a,Coraggio24a,DeGregorio24}, we have followed the approach that has been introduced by Suzuki and Okamoto \cite{Suzuki95}, which allows a construction of decay operators
\thetaeff~ which is consistent with the derivation of \heff~ (see the previous Sec. \ref{effham}).
In fact, it is grounded on the perturbative expansion of a vertex function \tbox, which plays the same role as the one of the \qbox~ within the derivation of \heff. 
A detailed description of such a procedure is reported in Refs. \cite{Suzuki95,Coraggio20c}, and in the following we sketch out briefly the procedure to derive effective SM decay operators \thetaeff~ by way of many-body perturbation theory.

The starting point is the perturbative calculation of two energy-dependent vertex functions:

\[
\hat{\Theta} (\epsilon) = P \Theta P + P \Theta Q
\frac{1}{\epsilon - Q H Q} Q H_1 P ~, \]
\[ \hat{\Theta} (\epsilon_1 ; \epsilon_2) = P H_1 Q
\frac{1}{\epsilon_1 - Q H Q} Q \Theta Q \frac{1}{\epsilon_2 - Q H Q} Q H_1 P ~,\]

\noindent
and of their derivatives calculated in $\epsilon=\epsilon_0$,
$\epsilon_0$ being the eigenvalue of the degenerate unperturbed
Hamiltonian $H_0$:

\[
\hat{\Theta}_m = \frac {1}{m!} \frac {d^m \hat{\Theta}
 (\epsilon)}{d \epsilon^m} \biggl|_{\epsilon=\epsilon_0} ~, \]
\[ \hat{\Theta}_{mn} =  \frac {1}{m! n!} \frac{d^m}{d \epsilon_1^m}
\frac{d^n}{d \epsilon_2^n}  \hat{\Theta} (\epsilon_1 ;\epsilon_2)
\biggl|_{\epsilon_1= \epsilon_0, \epsilon_2  = \epsilon_0} ~\]

Then, a series of operators $\chi_n$ is calculated:

\begin{eqnarray}
\chi_0 &=& (\hat{\Theta}_0 + h.c.)+ \hat{\Theta}_{00}~~,  \label{chi0} \\
\chi_1 &=& (\hat{\Theta}_1\hat{Q} + h.c.) + (\hat{\Theta}_{01}\hat{Q}
+ h.c.) ~~, \label{chi1} \\
\chi_2 &=& (\hat{\Theta}_1\hat{Q}_1 \hat{Q}+ h.c.) +
(\hat{\Theta}_{2}\hat{Q}\hat{Q} + h.c.) + \nonumber \\
~ & ~ & (\hat{\Theta}_{02}\hat{Q}\hat{Q} + h.c.)+  \hat{Q}
\hat{\Theta}_{11} \hat{Q}~~, \label{chi2} \\
&~~~& \cdots \nonumber
\end{eqnarray}

\noindent
where $\hat{Q} \equiv \hat{Q}(\epsilon)$.

At the end, \thetaeff~ is written in the following form:
\begin{equation}
\Theta_{\rm eff} = H_{\rm eff} \hat{Q}^{-1}  (\chi_0+ \chi_1 + \chi_2 +\cdots) ~,
\label{effopexp}
\end{equation}

\noindent
the $\chi_n$ series being arrested in our calculations at $n=2$, and the $\hat{\Theta}$ function expanded up to third order in perturbation theory.

In Refs. \cite{Coraggio18,Coraggio19a,Coraggio20a} we have also performed a study of the convergence of the $\chi_n$ series and of the perturbative properties of the \tbox, evidencing the robustness of the expansion of \thetaeff.

It should be pointed out that any effective single-body decay operator own also two- and many-body components, if the nuclear system is characterized by two or more valence nucleons.
We have included the leading order of the perturbative expansion of \tbox, namely the second-order two-body term, thanks also to the shell-model code KSHELL which allows to include also two-body components of the transition operators \cite{KSHELL}. 
In Ref. \cite{Coraggio25a}, the details of the calculation of this contribution to \thetaeff~ are reported, as well as a discussion about its impact on the calculation of the GT$^-$ strengths in $^{100}$Mo.

In this work, the decay operators $\Theta$ that are considered are the one-body electric-quadrupole $E2$ and magnetic $M1$ transition operators, GT-decay operator, as well as the one that rules the OMC. 

The latter will be introduced in the following section.

\subsection{ OMC  theory}
\label{operators}
We construct the operators which characterize the ordinary muon-capture process by following the formalism developed by Morita and Fujii~\cite{Morita60}, and in this section we briefly outline such a theoretical framework. 
Throughout this work, we adopt natural units ($\hbar=c=m_e=1$).

The transition rate from an initial state $|i\rangle$ with total angular momentum $J_i$ and energy $E_i$, to the final state $\langle f|$  with total angular momentum $J_f$ and energy $E_f$ is given by Fermi's golden rule
\begin{equation}
  \mathcal{W} = 2\pi \langle|\mathrm{M.E.}|^2\rangle_{\mathrm{av}} q^2 \frac{d q}{d E_f} \,,
\end{equation}
where $q$ is the momentum of the neutrino that is defined as

\begin{equation}
  q = (m_\mu - W_0) \left(1-\frac{m_\mu}{2(m_\mu + A m_N)} \right) \,,
\end{equation}
and the phase space factor is
\begin{equation}
  \frac{d q}{d E_f} = 1-\frac{q}{m_\mu + Am_N} \,.
 \end{equation}

 Here, $W_0$ is the energy difference between the final and initial nuclear states plus one electron mass, i.e. $W_0=E_f -E_i +m_e$, $A$ is the mass number, $m_N$ the average nucleon mass, $m_e$ and $m_\mu$ are the rest masses of electron and muon, respectively. 

The absolute square of the matrix element $\langle|\mathrm{M.E.}|^2\rangle_{\mathrm{av}}$, averaged over the initial and summed over the final substates, is defined in terms on nuclear matrix elements of OMC operators ($\mathfrak{M}_{vu}^{(\alpha)}$) as
\begin{eqnarray}
     \langle|\mathrm{M.E.}|^2\rangle_{\mathrm{av}} = \frac{1}{2}\frac{(2J_f+1)}{(2J_i+1)} \times\sum_{\alpha\beta}\sum_{\kappa u} C^{(\alpha)}C^{(\beta)} \nonumber \\ \times \left[\sum_{v}\mathfrak{M}_{vu}^{(\alpha)}(\kappa)\right]\left[\sum_{v^{\prime}}\mathfrak{M}_{v^{\prime}u}^{(\beta)}(\kappa)\right].
\end{eqnarray}
{
\renewcommand{\arraystretch}{1.5}
\begin{table*}[ht!] \centering
  \caption{Definition of $\Xi^{(\alpha)}$ in Eq.~\eqref{eqn:mes} for different OMC nuclear matrix elements (NMEs).
    In the last line, the $+$ and $-$ signs correspond to $\alpha=7$ and $8$, respectively.}
  \begin{ruledtabular}
  \begin{tabular}{l c l}

    \centering $\alpha$ & $C^{(\alpha)}$ & \multicolumn{1}{c}{$\Xi_{v u}^{(\alpha)}$} \\
    \hline
    $1$ & $C_{\mathrm{V}}$ & $\mathcal{Y}_{0 v u}^{M_f-M_i}\left(\hat{r}_s\right)\left[g_\kappa G_{\kappa^{\prime}} S_{0 v u}\left(\kappa, \kappa^{\prime}\right) -f_\kappa F_{\kappa^{\prime}} S_{0 v u}\left(-\kappa,-\kappa^{\prime}\right)\right] \delta_{v u}$ \\

    2 & $-C_{\mathrm{A}}$ & $\mathcal{Y}_{1 v u}^{M_f-M_i}\left(\hat{r}_s, \boldsymbol{\sigma}_s\right)\left[g_\kappa G_{\kappa^{\prime}} S_{1 v u}\left(\kappa, \kappa^{\prime}\right)-f_\kappa F_{\kappa^{\prime}} S_{1 v u}\left(-\kappa,-\kappa^{\prime}\right)\right]$ \\

    3 & $-C_{\mathrm{V}}/{m_N}$ & $i\left[f_\kappa G_{\kappa^{\prime}} S_{1 v u}\left(-\kappa, \kappa^{\prime}\right)+g_\kappa F_{\kappa^{\prime}} S_{1 v u}\left(\kappa,-\kappa^{\prime}\right)\right] \mathcal{Y}_{1 v u}^{M_f-M_i}\left(\hat{r}_s, \mathbf{p}_s\right)$ \\

    \multirow{2}{*}{4} & \multirow{2}{*}{$-\sqrt{3}C_{\mathrm{V}}/{2 m_N}$} & $\left(\sqrt{\frac{v+1}{2 v+3}} \mathcal{Y}_{0 v+1 u}^{M_f-M_i}\left(\hat{r}_s\right) \delta_{v+1 u} D_{+}-\sqrt{\frac{v}{2 v-1}} \mathcal{Y}_{0 v-1 u}^{M_f-M_i}\left(\hat{r}_s\right) \delta_{v-1 u} D_{-}\right)$ \\
     &  & $ \hfill \times \left[f_\kappa G_{\kappa^{\prime}} S_{1 v u}\left(-\kappa, \kappa^{\prime}\right)+g_\kappa F_{\kappa^{\prime}} S_{1 v u}\left(\kappa,-\kappa^{\prime}\right)\right]$ \\

    \multirow{3}{*}{5} & \multirow{3}{*}{ $-\sqrt{\frac{3}{2}} C_{\mathrm{V}}\left(1+\mu_p-\mu_n\right)/{m_N}$} & $\left(\sqrt{v+1} W(11 u v, 1 v+1) \times \mathcal{Y}_{1 v+1 u}^{M_f-M_i}\left(\hat{r}_s, \sigma_s\right) D_{+} \right.$ \\
     & & \qquad $\left.  -\sqrt{v} W(11 u v, 1 v-1) \mathcal{Y}_{1 v-1 u}^{M_f-M_i}\left(\hat{r}_s, \boldsymbol{\sigma}_s\right) D_{-}\right) $ \\
     & & \hfill $\times\left[f_\kappa G_{\kappa^{\prime}} S_{1 v u}\left(-\kappa, \kappa^{\prime}\right)+g_\kappa F_{\kappa^{\prime}} S_{1 v u}\left(\kappa,-\kappa^{\prime}\right)\right]$ \\

    6 & $C_{\mathrm{A}}/{m_N}$ & $i \mathcal{Y}_{0 v u}^{M_f-M_i}\left(\hat{r}_s\right)\left[f_\kappa G_{\kappa^{\prime}} S_{0 v u}\left(-\kappa, \kappa^{\prime}\right)+g_\kappa F_{\kappa^{\prime}} S_{0 v u}\left(\kappa,-\kappa^{\prime}\right)\right] \boldsymbol{\sigma}_s \cdot \mathbf{p}_s$ \\

    7 & $ -C_{\mathrm{A}}/{2\sqrt{3}m_N} $  & $\left(\sqrt{\frac{v+1}{2 v+1}} \mathcal{Y}_{1 v+1 u}^{M_f-M_i}\left(\hat{r}_s, \boldsymbol{\sigma}_s\right) D_{+}-\sqrt{\frac{v}{2 v+1}} \mathcal{Y}_{1 v-1 u}^{M_f-M_i}\left(\hat{r}_s, \boldsymbol{\sigma}_s\right) D_{-}\right)$ \\

    8 & $ C_{\mathrm{P}}/{2\sqrt{3} m_N}$ & \hfill $\times\left[f_\kappa G_{\kappa^{\prime}} S_{0 v u}\left(-\kappa, \kappa^{\prime}\right) \pm g_\kappa F_{\kappa^{\prime}} S_{0 v u}\left(\kappa,-\kappa^{\prime}\right)\right] \delta_{v u}$ \\

  \end{tabular}
  \end{ruledtabular}
  \label{tab:operators_cas}
\end{table*}
}
\noindent
In the above expression, the factor $1/2$ arises from the assumption that the muon occupies the lowest $1s_{\frac{1}{2}}$ state, $\kappa$ labels the quantum number of the neutrino, while $v$ and $u$ denote the total orbital angular momentum of the neutrino-muon system and the tensor rank of the capture operator, respectively. 
Additionally, $\mathfrak{M}_{vu}^{(\alpha)}$ are the reduced nuclear-matrix elements (NMEs) of the OMC transition operator $\Xi^{(\alpha)}$, and $C^{(\alpha)}$ are the corresponding coupling constants, both reported in Table \ref{tab:operators_cas}.

In a SM calculation, $\mathfrak{M}_{vu}^{(\alpha)}$ is expressed in terms of the single-particle matrix elements of the one-body decay operator (SPMEs) and the one-body transition densities (OBTDs), that can be obtained from the SM wave functions through the expression
\begin{equation}
    \mathfrak{M}_{vu}^{(\alpha)} ={\frac{1}{\hat{J_f}}}
    \sum_{\pi,\nu} \langle \nu || \Xi^{(\alpha)}_{vu}||\pi \rangle  \times
    {\rm OBTD}(\Psi_f,\Psi_i,\pi,\nu,u) ~,
    \label{eqn:mes}
\end{equation}
where $\hat{J_f}=\sqrt{2J_f +1}$, and $\pi(\nu)$ labels the proton (neutron) single-particle states. 
The explicit expression of the  operators $\Xi^{(\alpha)}$ in HO basis can be found in Ref.~\cite{Jenkins65}. 

In order to make explicit the expressions in Table \ref{tab:operators_cas}, here we list the quantities as reported:

\begin{itemize}
    \item [a)] The vector spherical harmonics $\mathcal{Y}_{kvu}^M$ are defined as
    \begin{equation}
    \begin{split}
      \mathcal{Y}_{0v u}^M(\hat{\bm{r}}) &= \sqrt{\frac{1}{4\pi}} Y_{v}^{M}(\hat{\bm{r}})\,,\\
      \mathcal{Y}_{1 v u}^M(\hat{\bm{r}}, \boldsymbol{\sigma}) &= \sum_m C_{1-m,v m+M}^{u M} Y_{v}^{m+M}(\hat{\bm{r}}) \mathcal{Y}_1^{-m}( \boldsymbol{\sigma})~~,
    \end{split}
    \end{equation}
    where the $C^{JM}_{j_1m_1j_2m_2}$ and $\mathcal{Y}_{l}^m$ (${Y}_{l}^m$) are the Clebsch-Gordan and the solid (spherical) harmonics, respectively.

    \item [b)] The radial wave functions of neutrino are
    \begin{equation}
        g_{\kappa}=\pi^{-1/2}j_l(qr) \,,\; f_\kappa=\pi^{-1/2}S_\kappa j_{\bar{l}}(qr)~~,
    \end{equation}
    where $j_l(qr)$ are the spherical Bessel functions and $S_\kappa$ represents the sign of $\kappa$. The total and orbital angular momentum, $j$ and $l$ are determined by $\kappa$ through the relation
    \begin{equation}
    \begin{split}
        l =\kappa \,, \; j=l-\frac12\quad &\mathrm{for}\quad\kappa>0 \,, \\
        l =-\kappa-1 \,, \; j=l+\frac12\quad &\mathrm{for}\quad\kappa<0 \,.
    \end{split}
    \end{equation}
    The subscripts $l$ and $\bar{l}$ denote the orbital angular momenta corresponding to $\kappa$ and $-\kappa$, respectively.
    
    \item [c)] The radial functions $F_{\kappa'}$ and $G_{\kappa'}$ are the {\it small} and {\it large} components of the bound state muon wave function obtained within the Bethe-Salpeter approximation \cite{Bethe59}. In the lowest $1s_{\frac{1}{2}}$ orbit, $\kappa'=-1$ and we have
    \begin{equation}
    \begin{split}
      G_{-1} &= \left(\frac{2Z}{a_0}\right)^{\frac{3}{2}}\sqrt{\frac{1+\gamma}{2\Gamma(2\gamma+1)}}\left(\frac{2Zr}{a_0}\right)^{\gamma-1}e^{-\frac{Zr}{a_0}},  \\
      F_{-1} &=-\sqrt{\frac{1-\gamma}{1+\gamma}}G_{-1}\,, \; \gamma = \sqrt{1-(\alpha Z)^2} \,.
    \end{split}
  \end{equation}
  
Here $a_0$ is the Bohr radius of the muonic atom (the reduced mass of the muon has been taken into account), $Z$ is the atomic number of the initial nucleus, and $\alpha$ is the fine structure constant. $\Gamma(z)$ represents the Gamma function.
    The wave function above is derived under the assumption of a point nucleus. To account for finite-size effects, an effective charge $Z_\mathrm{eff}$ can be introduced~\cite{Primakoff59,Ford62}, which generally reduces the capture rate by a factor for each nuclei. This factor is not included in the present work because it does not significantly affect the relative comparisons between different theoretical models or the trends across isotopes, which are the primary focus of this study.

    \item [d)] The geometric factor is defined as
    \begin{equation}
        S_{kvu}(\kappa, \kappa') = \sqrt{2}\hat{l}\hat{l'}\hat{j}\hat{j'} C_{l0,l'0}^{v0} \times \begin{Bmatrix}
            l & l' & v \\
            j & j' & u \\
            \frac12 & \frac12 & k \\
        \end{Bmatrix}\,,
    \end{equation}
    that becomes
    \begin{equation}
    S_{0vu}(\kappa,-1)=\sqrt{\frac{2j+1}{2l+1}}\delta_{lv},
    \end{equation}
    \begin{equation}
    S_{1vu}(\kappa,-1)=\sqrt{2(2j+1)} W\left( \frac{1}{2},1,j,l,\frac{1}{2},u\right)\delta_{lv},
    \end{equation}
    under the assumption $\kappa'=-1$, with $W(\cdots)$ denoting the Racah coefficient.
    
    \item [e)] The derivative operators in spherical harmonic basis can be expressed as
    \begin{equation}
        D_+ = \frac{d}{dr} - \frac{v}{r}\,, \quad D_- = \frac{d}{dr} + \frac{v+1}{r} \,.
    \end{equation}

    \item [f)] $\mu_p$ and $\mu_n$ are the proton and neutron magnetic moment and we used the value $\mu_p - \mu_n =3.706$ here.
\end{itemize}

The coupling constants $C^{(\alpha)}$ listed in the table can be expressed in terms of the Fermi interaction constant $G_F$ and the standard weak interaction couplings -- vector ($g_V$), axial-vector ($g_A$), and pseudoscalar ($g_P$) -- as
\begin{equation}
    C_V = g_V(q^2) G_V\,, \; C_{(A,P)} = -g_{(A,P)}(q^2) G_V
\end{equation}
where $G_V=G_F \cos\theta_C$ and $\theta_C$ is the Cabibbo mixing angle. The commonly used dipole parametrization for the nucleon form factors is adopted 
\begin{equation}
\begin{split}
    &g_V(q^2)=\frac{g_V}{\left(1+q^2/\Lambda_V^2\right)^2}, \\
    &g_A(q^2)=\frac{g_A}{\left(1+q^2/\Lambda_A^2\right)^2},
\end{split}
\end{equation}
where $g_V=1$, $g_A=1.2723$ and the cutoff parameters are $\Lambda_V=0.84~$GeV and $\Lambda_A=1~$GeV. We use the Goldberger-Treiman expression for the induced pseudoscalar coupling,
\begin{equation}
    g_P(q^2) = \frac{2m_\mu m_N}{m_\pi^2 + q^2} g_A(q^2).
\end{equation}

The structure of the OMC operator, that is basically composed by axial, vector, and pseudoscalar components, drives to an effective SM operator that cannot be reduced to a mere tuning of the coupling constants $g_A,g_V$, since the size of the renormalization may depend on the nature of each component of the operator, and is dependent upon the chosen SM configuration.
Besides that, we consider also a two-body component -- that is relevant for many-valence-nucleon systems -- of our effective SM operator (see previous section), that complicates the attempt to reduce its action in terms of a quenching factor of the ratio $g_A/g_V$ \cite{Coraggio25a}.

\section{Results}\label{results}
In this section we present the results of our RSM calculations. 

Since it is our intention to evidence the correlation between the quality of the reproduction of spectroscopic observables and those related to the OMC, we discuss separately these two analyses, by comparing theoretical results obtained with the two \heffs~ derived from the \vlwk~ and ChPT potentials, respectively, with available data.

\begin{figure}[H]
\begin{center}
\includegraphics[scale=0.31,angle=0]{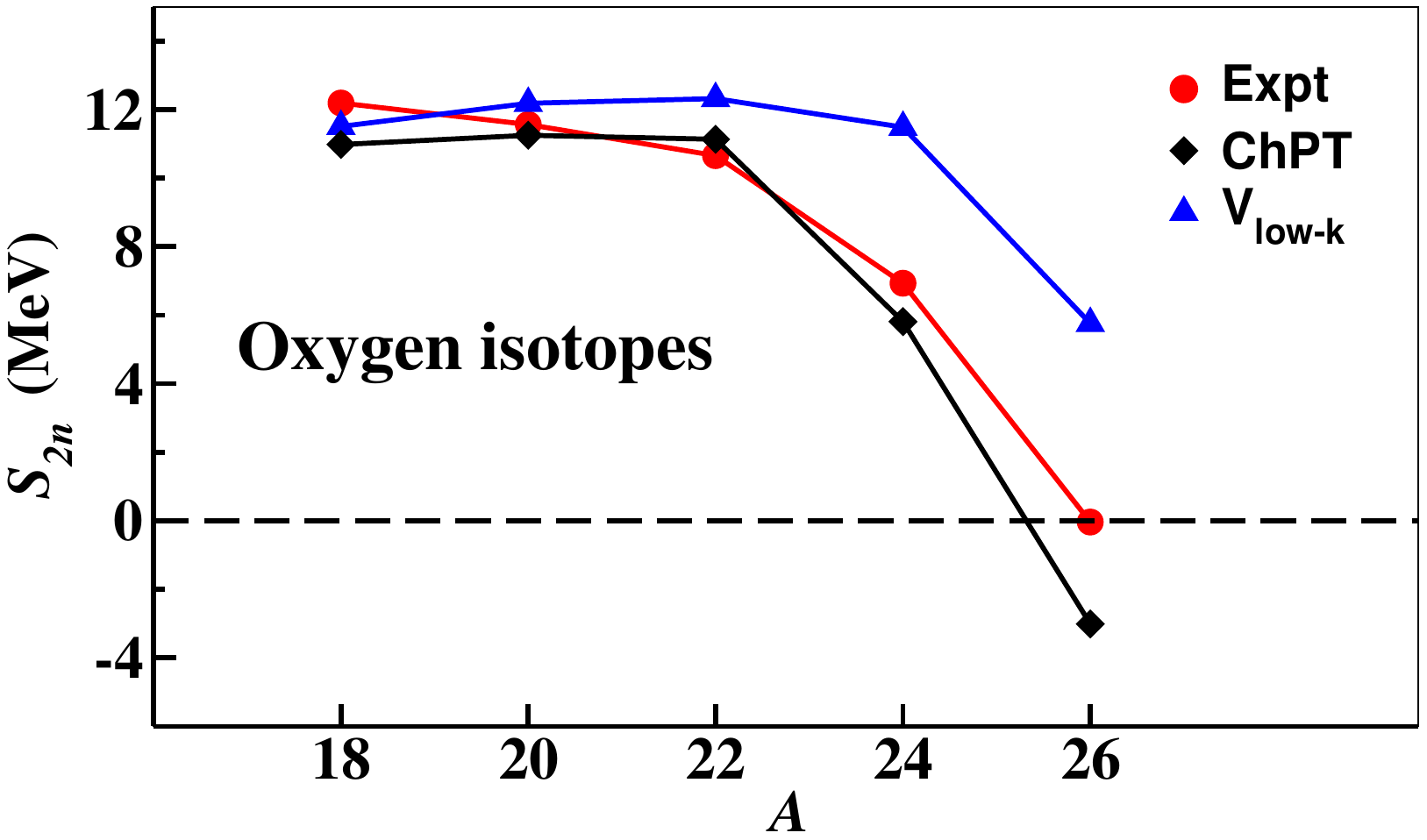}
\caption{Experimental and calculated two-neutron separation energies
  for even-mass oxygen isotopes from $A = 18$ to 26. Data are taken from
 \cite{Kondev21}, see text for details.}
\label{figS2n}
\end{center}
\end{figure}

\subsection{Spectroscopic properties}
First, we discuss the spectroscopic results, with a certain attention on the ability to reproduce the observed shell-evolution properties of nuclei in the $sd$-shell mass region.

As a matter of fact, even if oxygen isotopes are not the target of OMC studies, there are some relevant features of their spectroscopy as a function of the valence neutrons, that can characterize the shell-evolution of our \heffs.

We start showing in Fig. \ref{figS2n} our calculated two-neutron separation energies ($S_{2n}$) of even-mass isotopes up to $^{26}$O, compared with the experimental values \cite{Kondev21}. 

\begin{figure}[H]
\begin{center}
\includegraphics[scale=0.32,angle=0]{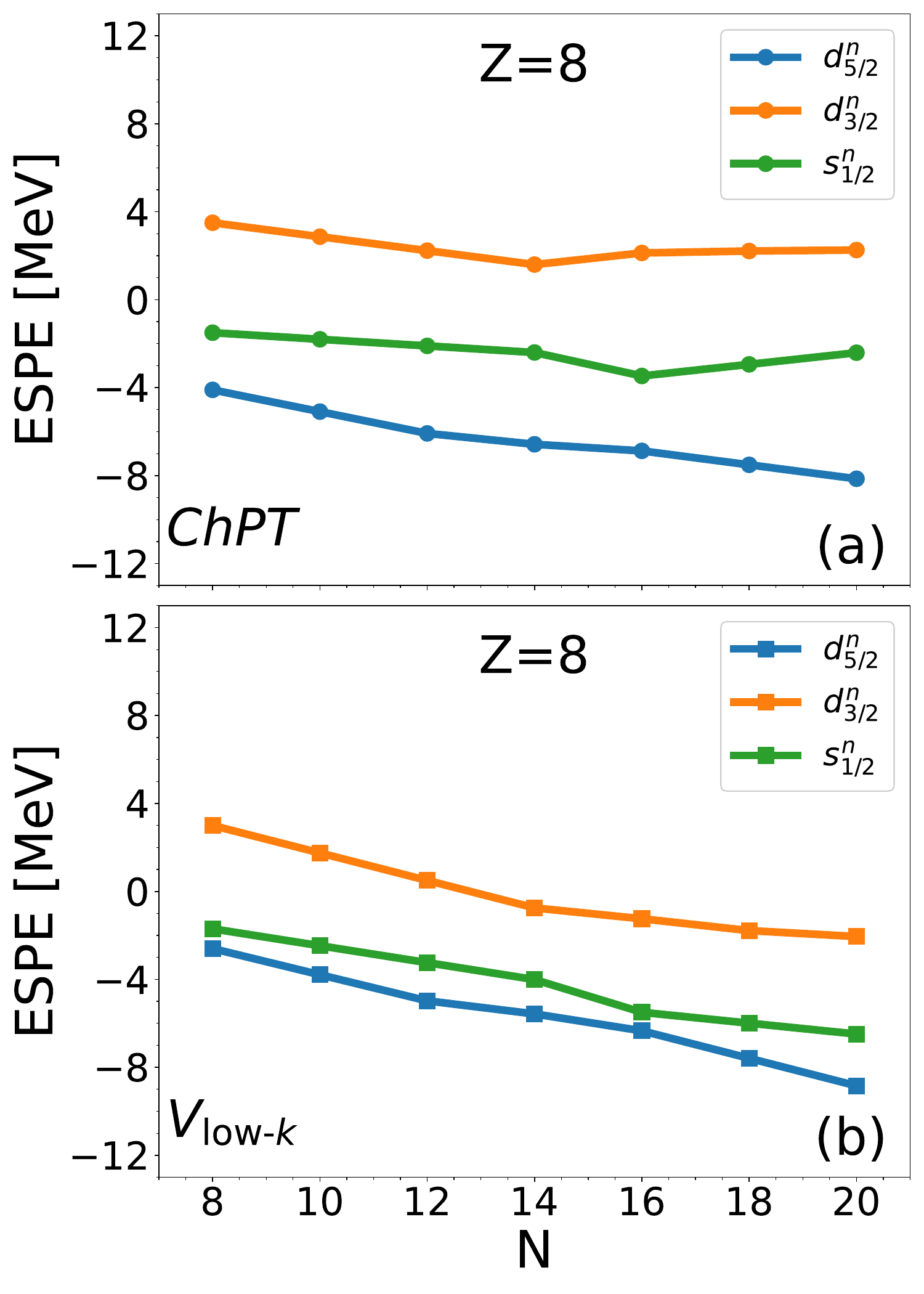}
\caption{Neutron ESPEs from ChPT (a) and \vlwk~ (b) \heffs~ for oxygen isotopes as a function of the neutron number.}
\label{espenn}
\end{center}
\end{figure}

As we have shown in \cite{Ma19}, we include enough intermediate states in the perturbative expansion of our \heffs~ to obtain convergent SP energy spacings as well as two-body matrix elements of the residual interaction (TBMEs), but they are not sufficient to obtain convergence for the ground-state (g.s.) energy of the nuclei with one-valence nucleon -- ($^{17}$F,$^{17}$O) -- with respect to the $^{16}$O core.
Then, the neutron SP energies -- that are reported in the Supplemental Material \cite{supplemental2025} together with proton SP energies and the TBMEs for both \vlwk~ and ChPT \heffs~ -- are shifted to reproduce the experimental g.s. energy of $^{17}$O with respect to $^{16}$O.
This allows to compare the different behavior of the $S_{2n}$, calculated with the two \heffs, but starting from the same value of the g.s. energy of $^{17}$O with respect the doubly-close $^{16}$O.

It is well known that the last bound oxygen isotope is $^{24}$O \cite{Kondo16}, then the ability to reproduce correctly the oxygen neutron dripline is a key feature of \heffs, reflecting the relevant aspects of their monopole components.

\begin{figure}[H]
\begin{center}
\includegraphics[scale=0.31,angle=0]{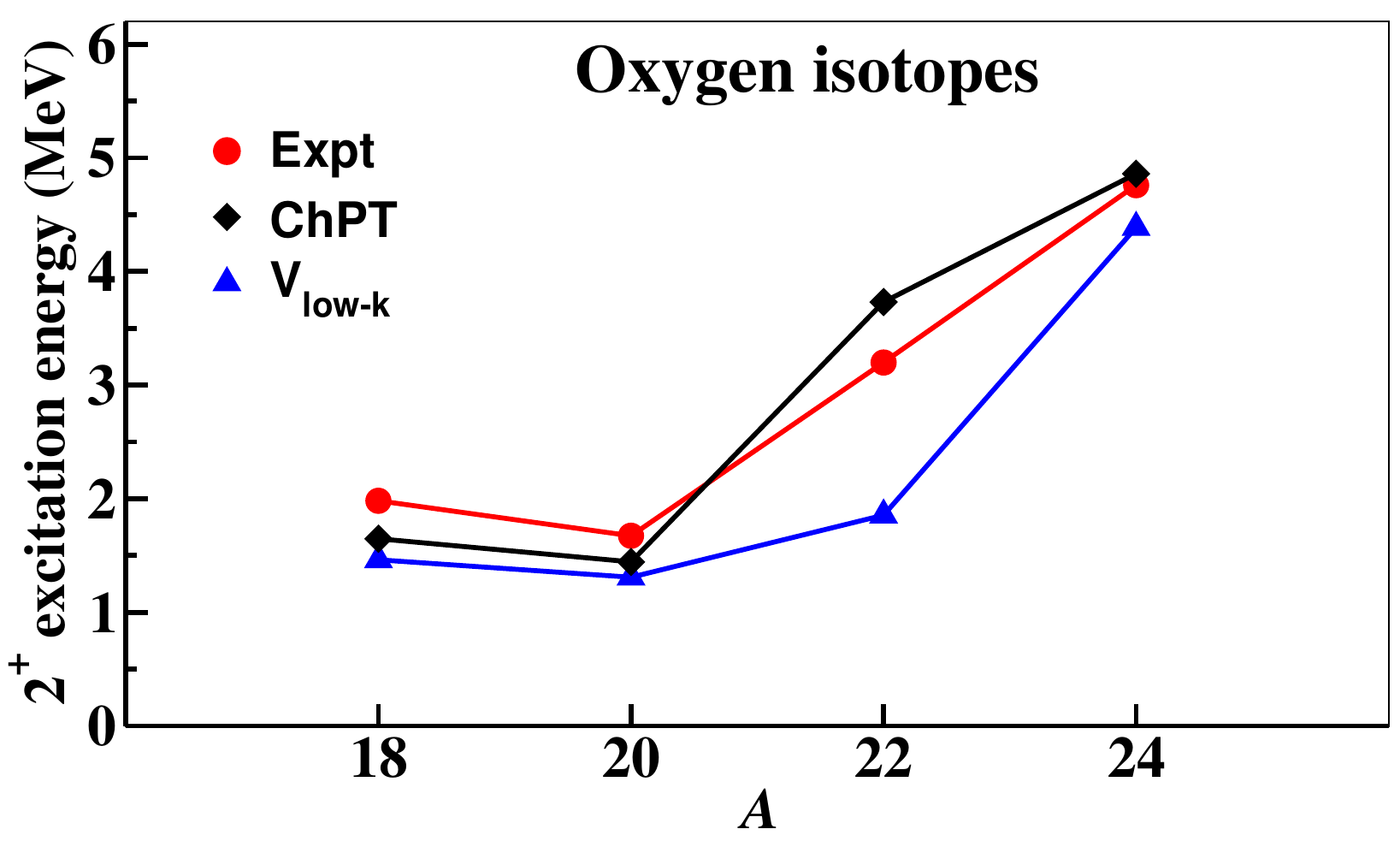}
\caption{Experimental and calculated excitation energies of the yrast
  $J^{\pi}=2^+$  states for even-mass oxygen isotopes from $A = 18$ to
  24. See text for details.}
\label{J2pOxygen}
\end{center}
\end{figure}

As can be seen in Fig. \ref{figS2n}, both \vlwk~ and ChPT \heffs~ reproduce the $S_{2n}$ experimental behavior up to $A=22$.
Then, the calculated results with ChPT \heff~ follow data and correctly an unbound $^{26}$O, while those obtained with \vlwk~ \heff~ depart from experiment, providing a bound $^{26}$O.
This different behavior traces back to the different energy spacings of the SP orbitals $0d_{5/2}$ and $1s_{1/2}$, as can also be seen in Fig. \ref{espenn} where the neutron effective SP energies (ESPEs) \cite{Utsuno99,Caurier05} are reported as a function of the neutron number $N$.
In particular, the fact that for $N=18$, when using ChPT \heff, the $0d_{3/2}$ SP orbital is unbound (i.e. $0d_{3/2}$ ESPE is positive) explains why the latter is able to reproduce correctly the neutron dripline.

Such an evolution of the neutron ESPEs determines also the behavior of the excitation energies $E^{1^{\rm st}}_{2^+}$ of yrast $J^{\pi}=2^+$ states of even-mass oxygen isotopes, that have been drawn in Fig. \ref{J2pOxygen}.
There, we may observe that there is a change in slope at $A=22$ of the values calculated with \vlwk~ \heff, with respect to the experimental one, with the drop being about 1 MeV, due to the absence of a gap between the $0d_{5/2}$ and $1s_{1/2}$ ESPEs.

\begin{center}
\begin{figure}[H]
\includegraphics[scale=0.2,angle=0]{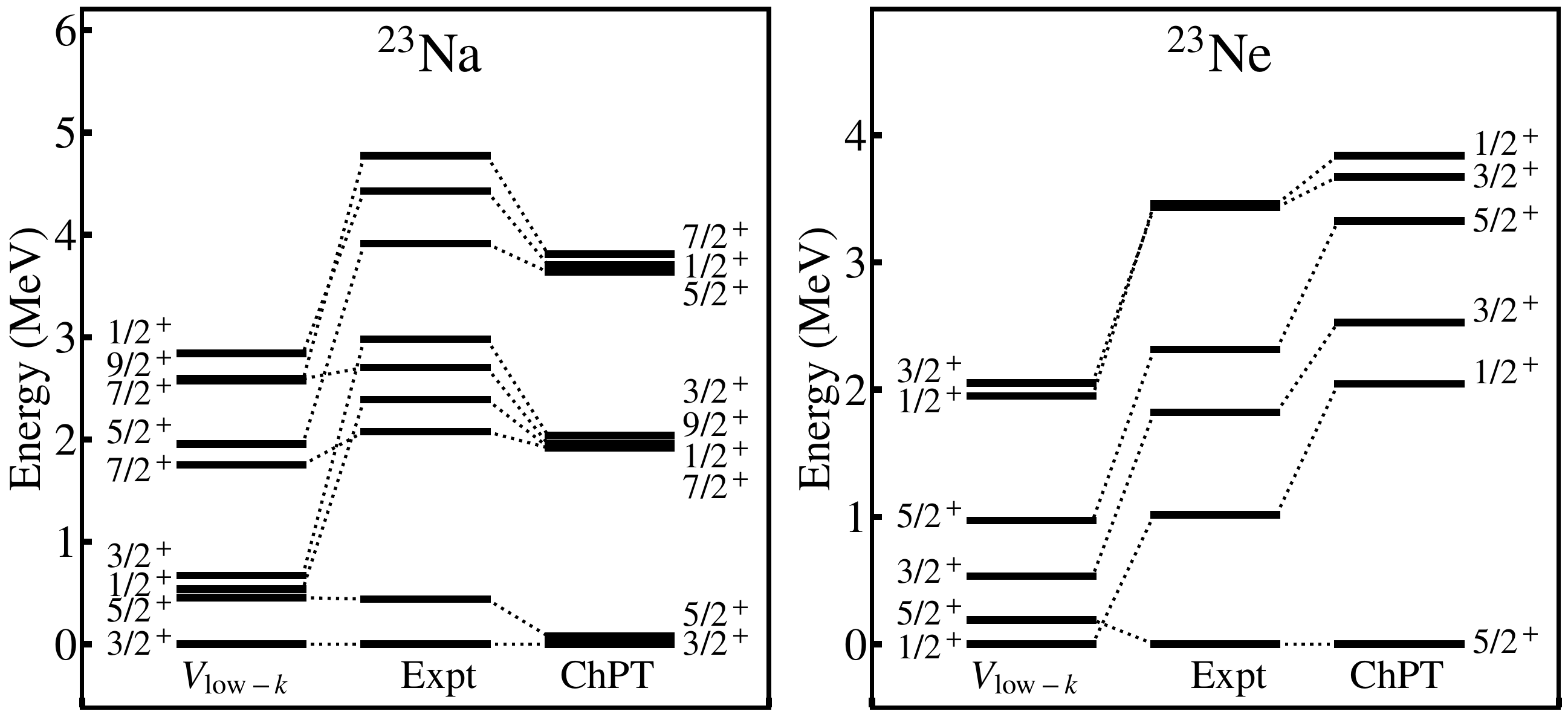}
\caption{Experimental and calculated low-energy spectra of $^{23}$Na and
  $^{23}$Ne. The calculated spectra are reported both with \vlwk~ and ChPT \heffs, experimental ones are taken from \cite{ensdf}.}
\label{23Na23Ne}
\end{figure}
\end{center}
\begin{center}
\begin{figure}[H]
\includegraphics[scale=0.2,angle=0]{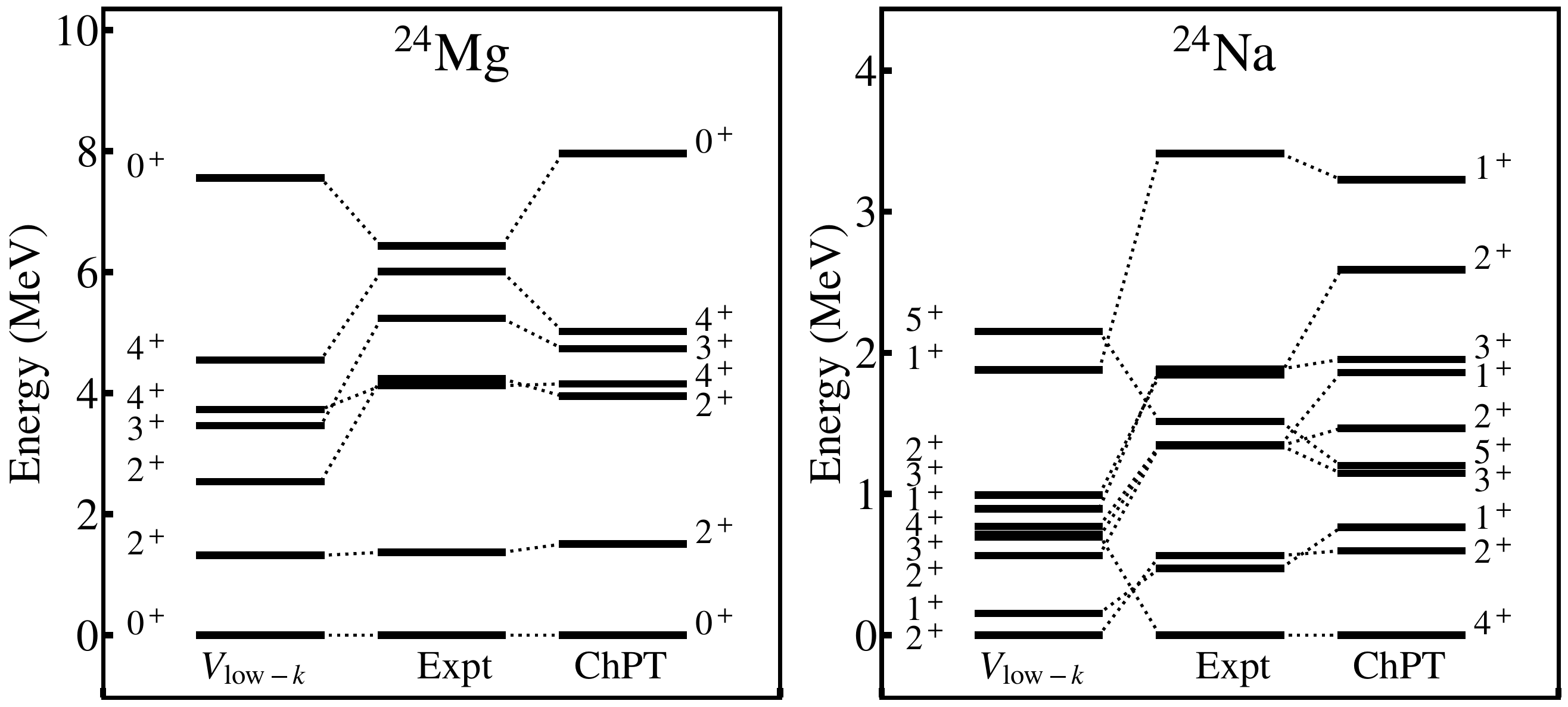}
\caption{Same as in Fig. \ref{23Na23Ne}, but for $^{24}$Mg and $^{24}$Na low-energy excitation spectra.}
\label{24Mg24Na}
\end{figure}
\end{center}
\begin{center}
\begin{figure}[H]
\includegraphics[scale=0.2,angle=0]{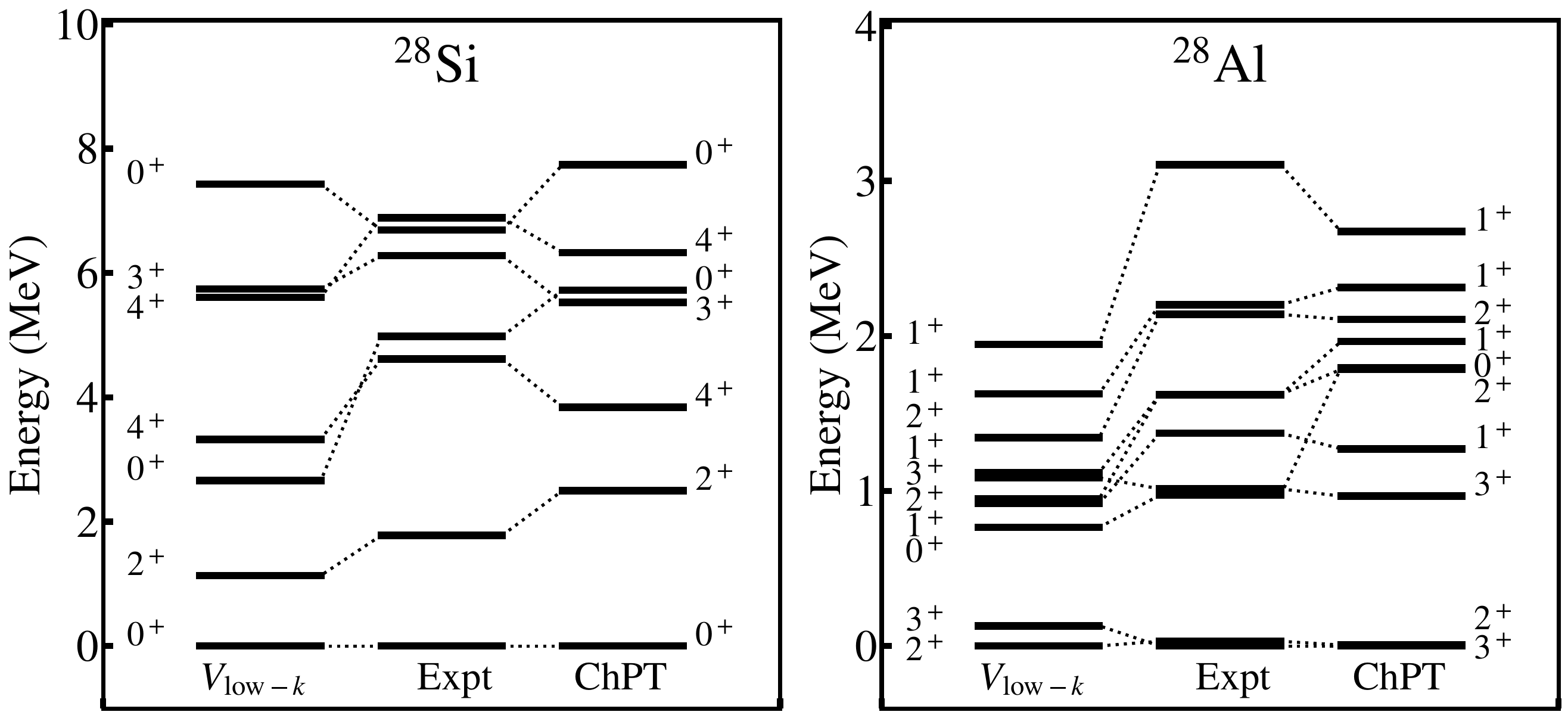}
\caption{Same as in Fig. \ref{23Na23Ne}, but for $^{28}$Si and $^{28}$Al low-energy excitation spectra.}
\label{28Si28Al}
\end{figure}
\end{center}
\begin{center}
\begin{figure}[H]
\includegraphics[scale=0.2,angle=0]{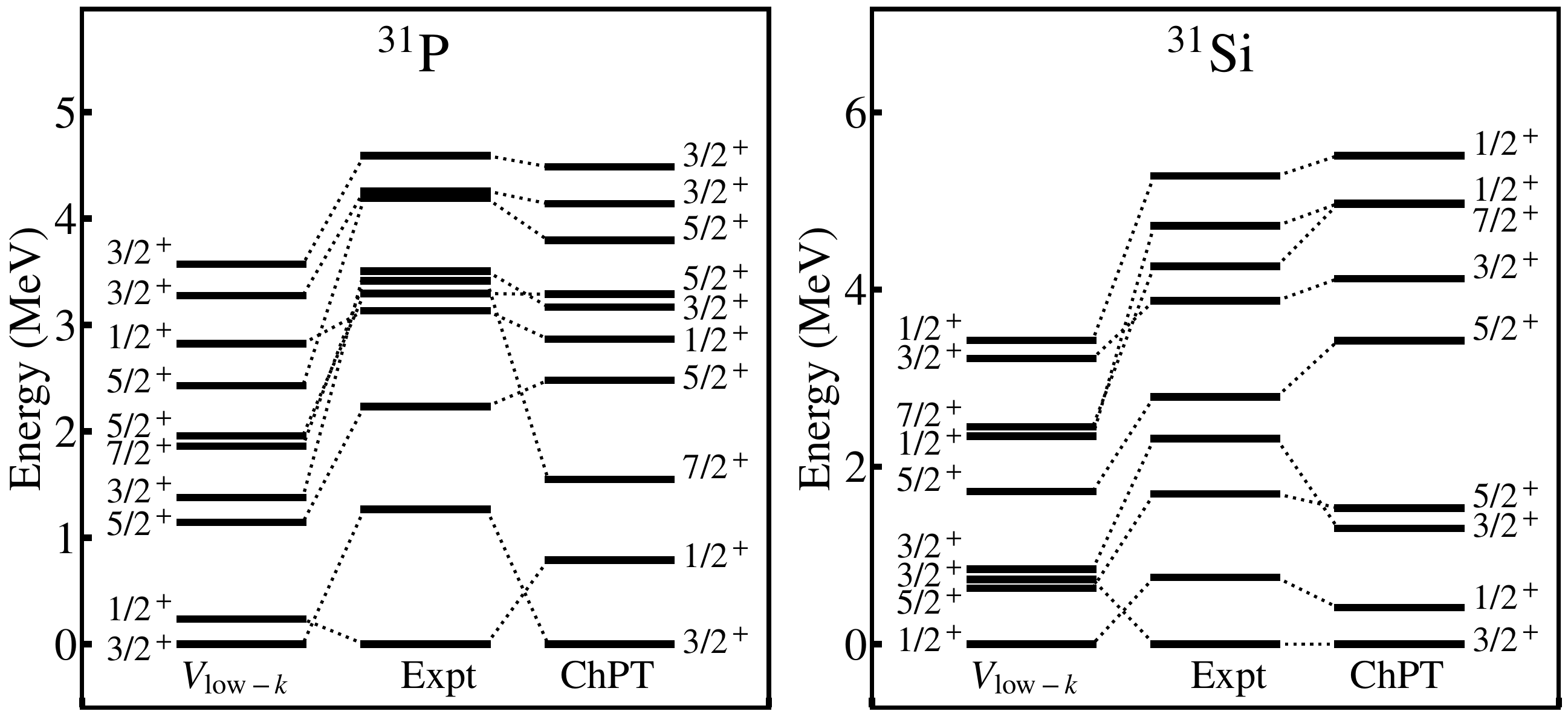}
\caption{Same as in Fig. \ref{23Na23Ne}, but for $^{31}$S and $^{31}$Si low-energy excitation spectra.}
\label{31S31Si}
\end{figure}
\end{center}
\begin{center}
\begin{figure}[H]
\includegraphics[scale=0.2,angle=0]{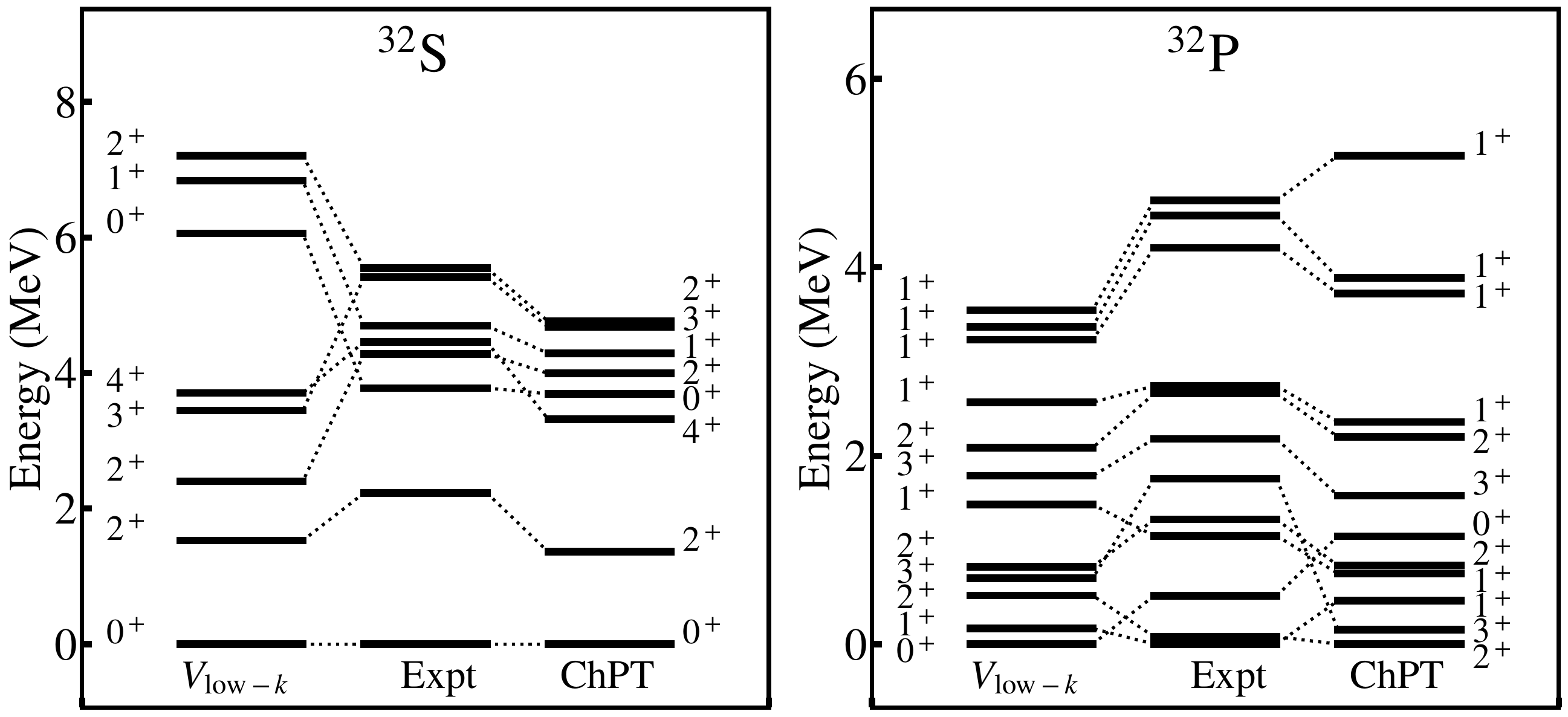}
\caption{Same as in Fig. \ref{23Na23Ne}, but for $^{32}$S and $^{32}$P low-energy excitation spectra.}
\label{32S32P}
\end{figure}
\end{center}

It should be pointed out that the ability to reproduce the experimental behavior of the binding energies of oxygen isotopes by accounting for a three-body component of the nuclear Hamiltonian was already evidenced in Ref. \cite{Otsuka10}, and a similar study has been performed for the evolution of $S_{2n}$ and  $E^{1^{\rm st}}_{2^+}$ for isotopic chains in $0f1p$ mass region \cite{Ma19,Coraggio20e,Coraggio21}.

In the following, we will consider the OMC partial capture rates in $^{23}$Na, $^{24}$Mg, $^{28}$Si, $^{31}$P, and $^{32}$S.
First, in Figs. \ref{23Na23Ne}-\ref{32S32P}, we show the low-energy excitation spectra of the nuclei involved in the OMC transitions under investigation, calculated with both ChPT and \vlwk~ \heffs~ and compared with the experimental counterparts.
For the odd-odd nuclei, all states with excitation energies below approximately $2$ MeV are included. For energies above $2$ MeV, only those states relevant to the muon capture transitions under investigation are shown, in order to enhance the clarity of the presentation.

From the inspection of Figs. \ref{23Na23Ne}-\ref{32S32P}, we observe that, in general, ChPT \heff~ provides a better reproduction of the low-energy spectra than the \vlwk~ one.
In particular, we point out that overall ChPT \heff~ performs better in reproducing the correct sequences of ground and first excited states in odd-even and odd-odd isotopes, the latter being a high appreciable feature for SM calculations.

\begin{figure}[H]
\begin{center}
\includegraphics[scale=0.32,angle=0]{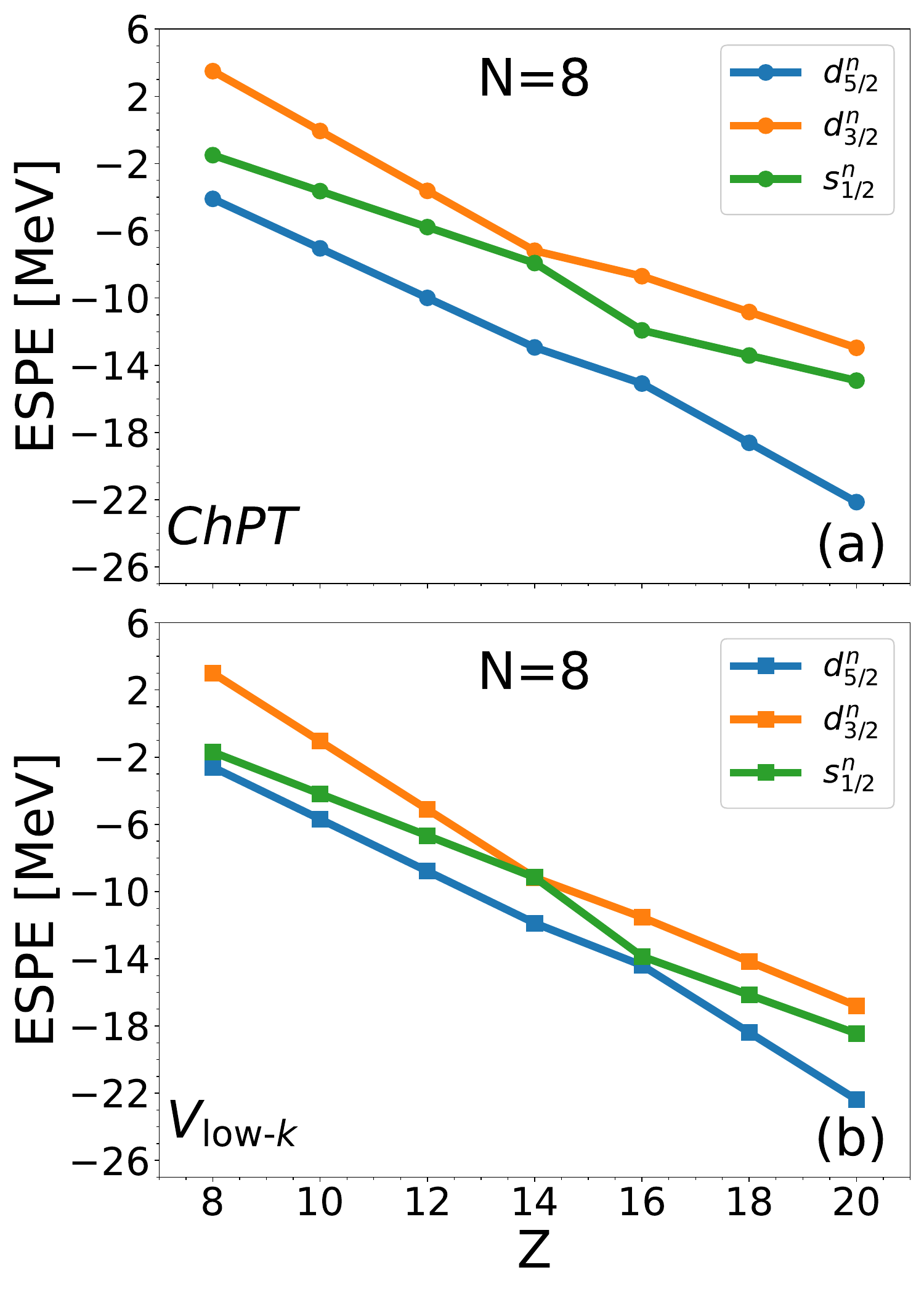}
\caption{Same as in Fig. \ref{espenn}, but for neutron ESPEs as a function of the proton number.}
\label{espepn}
\end{center}
\end{figure}

In general, \vlwk~ \heff~ provides a larger collectivity for the low-energy spectra, and this is a reflection of different proton-neutron monopole components of the \vlwk~ and ChPT \heffs.
This can be inferred by the behavior of the neutron ESPEs as a function of the atomic number $Z$ of $sd$-shell nuclei, as reported in Fig. \ref{espepn}.

As can be observed, the energy spacings of ESPEs from \vlwk~ \heff~ are more compressed than the ones obtained with ChPT \heff, driving a more relevant role as played by the higher-multipolarity components of the \heff.

The larger collectivity of the \vlwk~ \heff~ reflects also in the calculated values of the electric-quadrupole strengths ($B(E2)$) and moments -- that are reported in Tables \ref{tab:EM_ini}-\ref{tab:Qmu} (as well as the magnetic-dipole strengths $B(M1)$ and moments) and compared with the available data \cite{ensdf} --, as obtained by employing the two different \heffs.

\begin{table*}[ht]
\centering
\scriptsize
\caption{$B(E2)$ (in $e^2\mathrm{fm}^4$) and $B(M1)$ (in $\mu_N^2$) calculated and experimental values.}
\label{tab:EM_mirrors}
\begin{ruledtabular}
\begin{tabular}{l l c r r r | l l c r r r}
\multicolumn{6}{c|}{\textbf{Initial nuclei}} & \multicolumn{6}{c}{\textbf{Final nuclei}} \\
Nucleus & ($J_i\to J_f$) & E2/M1 & \vlwk & ChPT & Expt &
Nucleus & ($J_i\to J_f$) & E2/M1 & \vlwk & ChPT & Expt \\
\hline
${}^{23}\mathrm{Na}$ & $(5/2^+ \to 3/2^+)$ & E2 & 95   & 71  & $124\pm23$  & ${}^{23}\mathrm{Ne}$ & $(1/2^+ \to 5/2^+)$ & E2 & 1   & 7  & $3\pm0.2$ \\
${}^{23}\mathrm{Na}$ & $(7/2^+ \to 5/2^+)$ & E2 & 60   & 31  & $57\pm9$    &                     &                       &    &     &    &           \\
${}^{23}\mathrm{Na}$ & $(7/2^+ \to 3/2^+)$ & E2 & 34   & 23  & $47\pm6$    &                     &                       &    &     &    &           \\
${}^{23}\mathrm{Na}$ & $(1/2^+ \to 5/2^+)$ & E2 & 5    & 18  & $11\pm3$    &                     &                       &    &     &    &           \\
${}^{23}\mathrm{Na}$ & $(9/2^+ \to 5/2^+)$ & E2 & 50   & 37  & $68\pm5$    &                     &                       &    &     &    &           \\
${}^{23}\mathrm{Na}$ & $(9/2^+ \to 7/2^+)$ & E2 & 29   & 31  & $233\pm12$  &                     &                       &    &     &    &           \\
${}^{23}\mathrm{Na}$ & $(5/2^+ \to 3/2^+)$ & M1 & 0.27 &0.19 & $0.40\pm0.03$ &                    &                       &    &     &    &           \\
${}^{23}\mathrm{Na}$ & $(7/2^+ \to 5/2^+)$ & M1 & 0.33 &0.09 & $0.29\pm0.03$ &                    &                       &    &     &    &           \\
${}^{23}\mathrm{Na}$ & $(9/2^+ \to 7/2^+)$ & M1 & 0.22 &0.33 & $0.68\pm0.05$ &                    &                       &    &     &    &           \\
\hline
${}^{24}\mathrm{Mg}$ & $(2^+ \to 0^+)$     & E2 & 74   & 52  & $87\pm2$    & ${}^{24}\mathrm{Na}$ & $(2^+ \to 4^+)$        & E2 & 10  & 16 & $10\pm2$ \\
${}^{24}\mathrm{Mg}$ & $(4^+ \to 2^+)$     & E2 & 98   & 73  & $147\pm14$  & ${}^{24}\mathrm{Na}$ & $(2^+_2 \to 4^+)$      & E2 & 1   & 3  & $2\pm1$  \\
${}^{24}\mathrm{Mg}$ & $(2^+_2 \to 2^+_1)$ & E2 & 38   & 9   & $14\pm1$    & ${}^{24}\mathrm{Na}$ & $(5^+ \to 4^+)$        & E2 & 73  & 39 & $66\pm45$\\
${}^{24}\mathrm{Mg}$ & $(2^+_2 \to 0^+)$   & E2 & 4    & 6   & $7\pm1$     & ${}^{24}\mathrm{Na}$ & $(4^+_2 \to 2^+_2)$    & M1 & 0.07&0.3 &$0.4\pm0.1$\\
${}^{24}\mathrm{Mg}$ & $(3^+ \to 2^+)$     & E2 & 7    & 8   & $8\pm1$     &                     &                       &    &     &    &           \\
${}^{24}\mathrm{Mg}$ & $(4^+_2 \to 2^+_2)$ & E2 & 28   & 26  & $61\pm5$    &                     &                       &    &     &    &           \\
\hline
${}^{28}\mathrm{Si}$ & $(2^+ \to 0^+)$     & E2 & 97   & 51  & $54\pm2$    & ${}^{28}\mathrm{Al}$ & $(0^+ \to 2^+)$        & E2 & 2   & 2  & $23\pm2$ \\
${}^{28}\mathrm{Si}$ & $(4^+ \to 2^+)$     & E2 & 130  & 49  & $67\pm7$    & ${}^{28}\mathrm{Al}$ & $(3^+_2 \to 2^+)$      & E2 & 61  &26  & $60\pm50$\\
${}^{28}\mathrm{Si}$ & $(0^+_2 \to 2^+_1)$ & E2 & 11   & 74  & $39\pm2$    & ${}^{28}\mathrm{Al}$ & $(1^+ \to 2^+)$        & E2 & 0.2 & 7  & $6\pm5$  \\
${}^{28}\mathrm{Si}$ & $(2^+_2 \to 0^+_2)$ & E2 & 102  & 20  & $3\pm2$     & ${}^{28}\mathrm{Al}$ & $(1^+ \to 3^+)$        & E2 & 4   &0.03& $45\pm25$\\
                     &                      &    &      &     &             & ${}^{28}\mathrm{Al}$ & $(2^+ \to 3^+)$        & M1 & 0.03&0.03& $0.66\pm0.01$\\
                     &                      &    &      &     &             & ${}^{28}\mathrm{Al}$ & $(3^+_2 \to 2^+)$      & M1 &0.014&0.26& $0.25\pm0.05$\\
                     &                      &    &      &     &             & ${}^{28}\mathrm{Al}$ & $(1^+ \to 0^+)$        & M1 &0.0002&0.007& $0.97\pm0.21$\\
\hline
${}^{31}\mathrm{P}$  & $(3/2^+ \to 1/2^+)$ & E2 & 34   & 31  & $21\pm5$    & ${}^{31}\mathrm{Si}$  & $(5/2^+ \to 1/2^+)$    & E2 & 55  & 14 & $14\pm6$ \\
${}^{31}\mathrm{P}$  & $(5/2^+ \to 1/2^+)$ & E2 & 13   & 23  & $39\pm3$    & ${}^{31}\mathrm{Si}$  & $(5/2^+ \to 3/2^+)$    & E2 & 33  & 27 & $67\pm23$\\
${}^{31}\mathrm{P}$  & $(5/2^+_2 \to 5/2^+_1)$ & E2 & 16 & 1 & $145\pm58$ & ${}^{31}\mathrm{Si}$  & $(3/2^+_2 \to 1/2^+)$    & E2 & 49  & 18 & $272\pm197$\\
${}^{31}\mathrm{P}$  & $(7/2^+ \to 3/2^+)$ & E2 & 71   & 33  & $57\pm22$   & ${}^{31}\mathrm{Si}$  & $(3/2^+_2 \to 5/2^+)$    & M1 & 0.10 &0.15& $0.43\pm0.30$\\
${}^{31}\mathrm{P}$  & $(3/2^+ \to 1/2^+)$ & M1 &0.092 &0.020& $0.036\pm0.002$ &                  &                       &    &     &    &           \\
${}^{31}\mathrm{P}$  & $(5/2^+_2 \to 5/2^+_1)$ & M1 &0.098 &0.015& $0.070\pm0.018$ &               &                       &    &     &    &           \\
${}^{31}\mathrm{P}$  & $(5/2^+_2 \to 3/2^+)$ & M1 &0.256 &0.009& $0.036\pm0.009$ &                &                       &    &     &    &           \\
\hline
${}^{32}\mathrm{S}$  & $(0^+_2 \to 2^+)$   & E2 & 1    & 2   & $71\pm7$    & ${}^{32}\mathrm{P}$   & $(1^+_2 \to 2^+)$      & E2 & 12  &22  & $18\pm18$\\
${}^{32}\mathrm{S}$  & $(2^+_2 \to 2^+_1)$ & E2 & 108  & 5   & $48\pm5$    & ${}^{32}\mathrm{P}$   & $(3^+ \to 2^+)$        & E2 & 0.5 &33  & $36\pm5$ \\
${}^{32}\mathrm{S}$  & $(4^+ \to 2^+)$     & E2 & 75   & 54  & $84\pm18$   & ${}^{32}\mathrm{P}$   & $(3^+_2 \to 1^+)$      & E2 & 2   &19  & $19\pm4$ \\
${}^{32}\mathrm{S}$  & $(3^+ \to 2^+)$     & E2 & 1    & 0.1 & $10\pm2$    & ${}^{32}\mathrm{P}$   & $(2^+ \to 1^+)$        & M1 & 0   &0.03& $0.30\pm0.01$\\
${}^{32}\mathrm{S}$  & $(1^+ \to 2^+)$     & M1 &0.002 &0.001& $0.006\pm0.001$ & ${}^{32}\mathrm{P}$ & $(1^+_2 \to 0^+)$      & M1 &0.0009&0.03& $0.42\pm0.03$\\
\end{tabular}
\end{ruledtabular}
\end{table*}

\begin{table}[ht]
\caption{Electric-quadrupole ($Q$ in barn) and magnetic-dipole moments ($\mu$ in $\mu_N$).}
\label{tab:Qmu}
\begin{ruledtabular}
\begin{tabular}{cccccc}
    Nucleus & $J^\pi$ & {$Q/\mu$} & \vlwk & ChPT & Expt \\ \hline
\multirow{2}{*}{${}^{23}$Na} 
    & $3/2^+$  & $Q$ & $0.102$ &$0.079$ & $+0.104(1)$ \\ \cline{2-6}
    & $3/2^+$  & $\mu$ & 1.802 &$1.626$ & $-2.21750(3)$ \\ \hline
\multirow{2}{*}{${}^{23}$Ne} 
    & $5/2^+$  & $Q$ & $0.142$ &$0.112$ & $+0.145(13)$ \\ \cline{2-6}
    & $5/2^+$  & $\mu$ & $-0.818$ &$-0.753$ & $-1.0794$ \\ \hline
\multirow{5}{*}{${}^{24}$Mg} 
    & $2^+_1$  & $Q$ & $-0.15$ & $-0.14$ & $-0.29(3)$  \\ \cline{2-6}
    & $2^+_1$  & $\mu$ & $1.02$ & $0.94$  & $+1.08(3)$ \\ 
    & $2^+_2$  & $\mu$ & $1.04$ & $0.95$  & $+1.3(4)$ \\
    & $4^+_1$  & $\mu$ & $2.06$ & $1.9$  & $+1.7(12)$ \\ 
    & $4^+_2$  & $\mu$ & $2.07$ & $1.9$  & $+2.1(16)$ \\\hline
\multirow{2}{*}{${}^{24}$Na} 
    & $1^+$  & $\mu$ & $0.28$ & $-1.26$  & $-1.931(3)$ \\
    & $4^+$  & $\mu$ & $1.45$ & $1.48$  & $+1.6903(8)$ \\ \hline
\multirow{2}{*}{${}^{28}$Si} 
    & $2^+$  & $Q$ & $0.20$ & $0.14$ & $+0.16(3)$  \\ \cline{2-6}
    & $2^+$  & $\mu$ & $1.01$ & $0.93$  & $+1.1(2)$ \\ \hline
\multirow{3}{*}{${}^{28}$Al} 
    & $3^+$  & $Q$ & $0.175$ & $0.115$ & $0.172(12)$  \\ \cline{2-6}
    & $2^+$  & $\mu$ & $0.16$ & $2.3$  & $+4.0(4)$ \\ 
    & $3^+$  & $\mu$ & $-0.72$ & $2.83$  & $3.241(5)$ \\ \hline
\multirow{3}{*}{${}^{31}$P} 
    & $1/2^+$  & $\mu$ & $0.18$ &$0.46$ & $+1.130925(5)$ \\
    & $3/2^+$  & $\mu$ & $0.33$ &$0.35$ & $+0.30(8)$ \\
    & $5/2^+$  & $\mu$ & $0.71$ & $1.1$ & $+2.8(6)$ \\ \hline
\multirow{3}{*}{${}^{32}$S} 
    & $2^+$  & $Q$ & $0.08$ & $-0.14$ & $-0.16(2)$  \\ \cline{2-6}
    & $2^+$  & $\mu$ & $1.04$ & $0.9$  & $+0.9(2)$ \\ 
    & $4^+$  & $\mu$ & $2.04$ & $1.7$  & $+1.6(6)$ \\ \hline
\multirow{1}{*}{${}^{32}$P} 
    & $1^+$  & $\mu$ & $-0.81$ & $0.50$  & $-0.2528(2)$ \\ 
\end{tabular}
\end{ruledtabular}
\end{table}

As can be seen, the calculated $B(E2)$s and electric-quadrupole moments are larger when employing \vlwk~ \heff~ than those with ChPT \heff.

It should be pointed out that we obtain an overall good agreement with experimental strengths, both for $E2$ and $M1$ transitions, by employing effective SM Hamiltonians and transition operators derived from the realistic \vlwk~ and ChPT potentials.

 In order to test the quality of our nuclear wave functions and effective decay operators, we complete our study of the decay properties of the low-energy states of the nuclei under consideration by comparing the calculated $\log ft$ of $\beta$ decay with available data, as reported in Table \ref{tab:logft}.

Again, the quality of the reproduction of the experimental values is quite remarkable, and testifies the ability of RSM to provide reliable effective SM Hamiltonians and decay operators.

\begin{table}[h]
\caption{$\log ft$ value for $\beta$ decay.}
\label{tab:logft}
\begin{ruledtabular}
\begin{tabular}{ccccc}
\multicolumn{1}{c}{\multirow{2}{*}{Decay}} & \multirow{2}{*}{($J_i \to J_f$)} & \multicolumn{3}{c}{$\log ft$} \\ \cline{3-5} 
    \multicolumn{1}{c}{}   &   & \vlwk &   ChPT & Expt  \\ \hline
    \multirow{3}{*}{${}^{23}$Ne $\rightarrow$ ${}^{23}$Na }   
    & $(5/2^+ \to 3/2^+)$  & $5.43$ & {$5.30$}  & {$5.27$}  \\
    & $(5/2^+ \to 5/2^+)$  & $5.55$ & {$5.32$}  & {$5.38$}  \\
    & $(5/2^+ \to 7/2^+)$  & $7.22$ & {$5.69$}  & {$5.82$}  \\ \hline
\multirow{2}{*}{${}^{24}$Na $\rightarrow$ ${}^{24}$Mg}                          
    & $(4^+ \to 4^+)$  & $6.71$ & {$5.45$}  & {$6.12$}  \\
    & $(4^+ \to 3^+)$  & $5.98$ & $6.58$   &  $6.66$   \\ \hline
\multirow{1}{*}{${}^{28}$Al $\rightarrow$ ${}^{28}$Si}    
    & $(3^+ \to 2^+)$  & $6.93$ & {$7.23$}  & {$4.86$}  \\ \hline
\multirow{2}{*}{${}^{31}$Si $\rightarrow$ ${}^{31}$P}  
    & $(3/2^+ \to 3/2^+)$ & $6.59$ & {$5.28$} & {$5.75$}  \\
    & $(3/2^+ \to 1/2^+)$ & $7.17$ & {$7.21$}  & {$5.53$}  \\ \hline
\multirow{1}{*}{${}^{32}$P $\rightarrow$ ${}^{32}$S}    
    & $(1^+ \to 0^+)$ & $7.95$  & {$6.03$}  & {$7.90$} \\
\end{tabular}
\end{ruledtabular}
\end{table}

\subsection{OMC partial rates}

\begin{figure}[h!]
    \centering
    \includegraphics[scale=0.2]{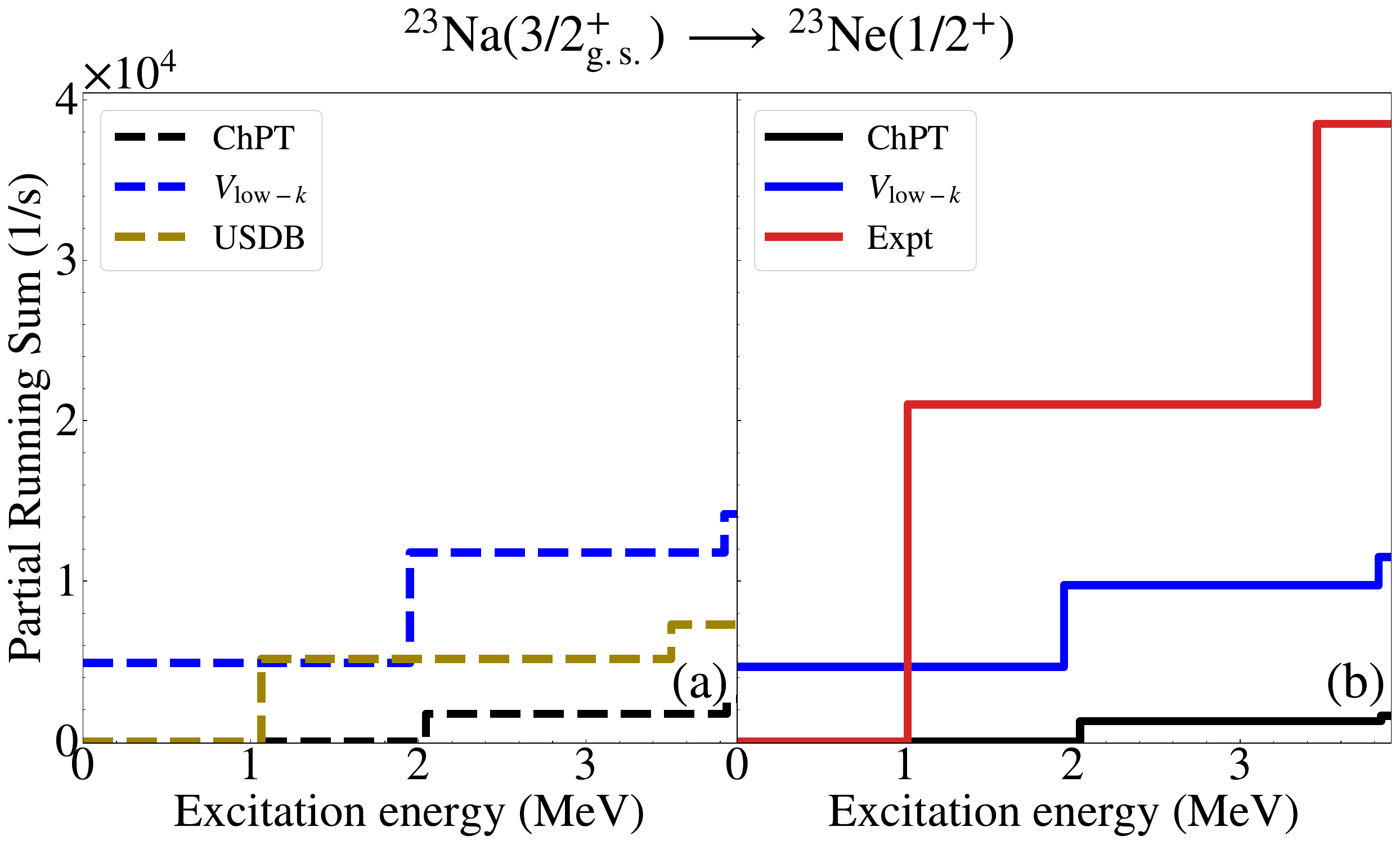}
    \caption{Capture rate from the ground state of ${}^{23}$Na to the ${}^{23}$Ne$(1/2^+)$ final states. In panel (a) we compare the results obtained with \vlwk, ChPT, and USDB \heffs, and employing the bare OMC transition operator. In panel (b) we compare the results with \vlwk~ and ChPT \heffs, employing their respective effective SM operators, and compared with the available data (see text for details).}
    \label{fig:rsum_Na_12}
\end{figure}

\begin{figure}[h!]
    \centering
    \includegraphics[scale=0.2]{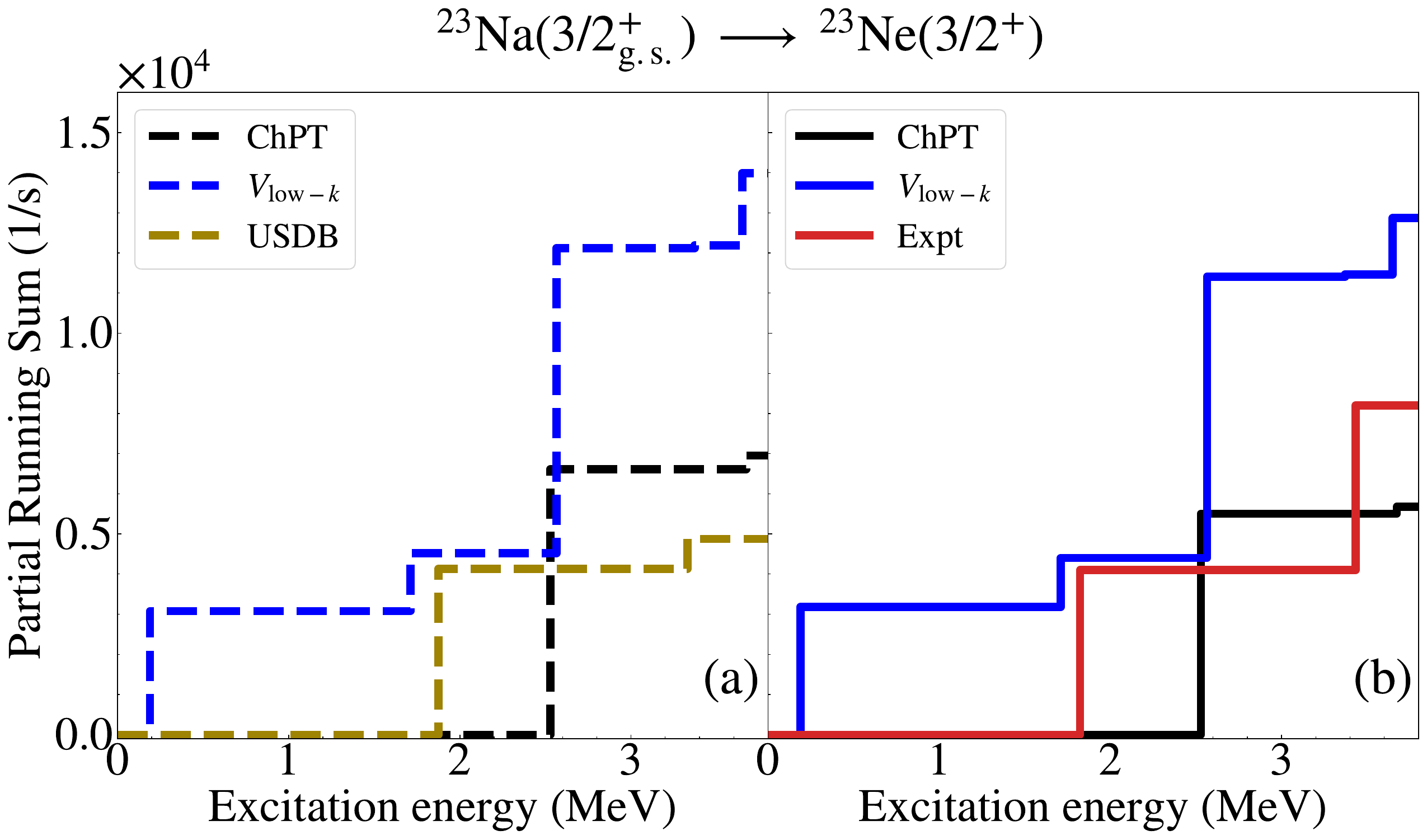}
    \caption{Same as in Fig. \ref{fig:rsum_Na_12}, but for the capture rate from the ground state of ${}^{23}$Na to the ${}^{23}$Ne$(3/2^+)$ final states.}
    \label{fig:rsum_Na_32}
\end{figure}

\begin{figure}[h!]
    \centering
    \includegraphics[scale=0.2]{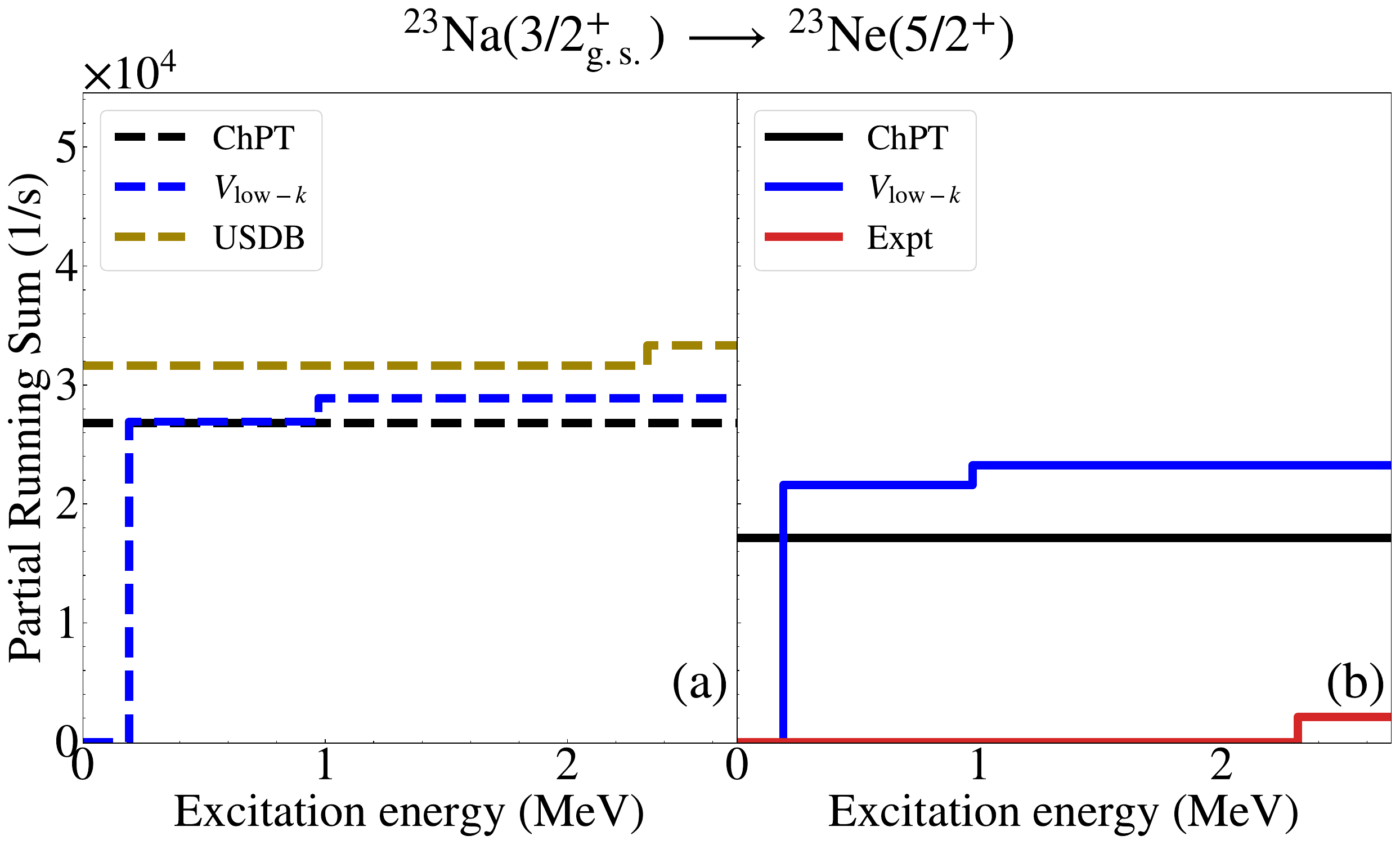}
    \caption{Same as in Fig. \ref{fig:rsum_Na_12}, but for the capture rate from the ground state of ${}^{23}$Na to the ${}^{23}$Ne$(5/2^+)$ final states.}
    \label{fig:rsum_Na_52}
\end{figure}

\begin{figure}[h!]
    \centering
    \includegraphics[scale=0.2]{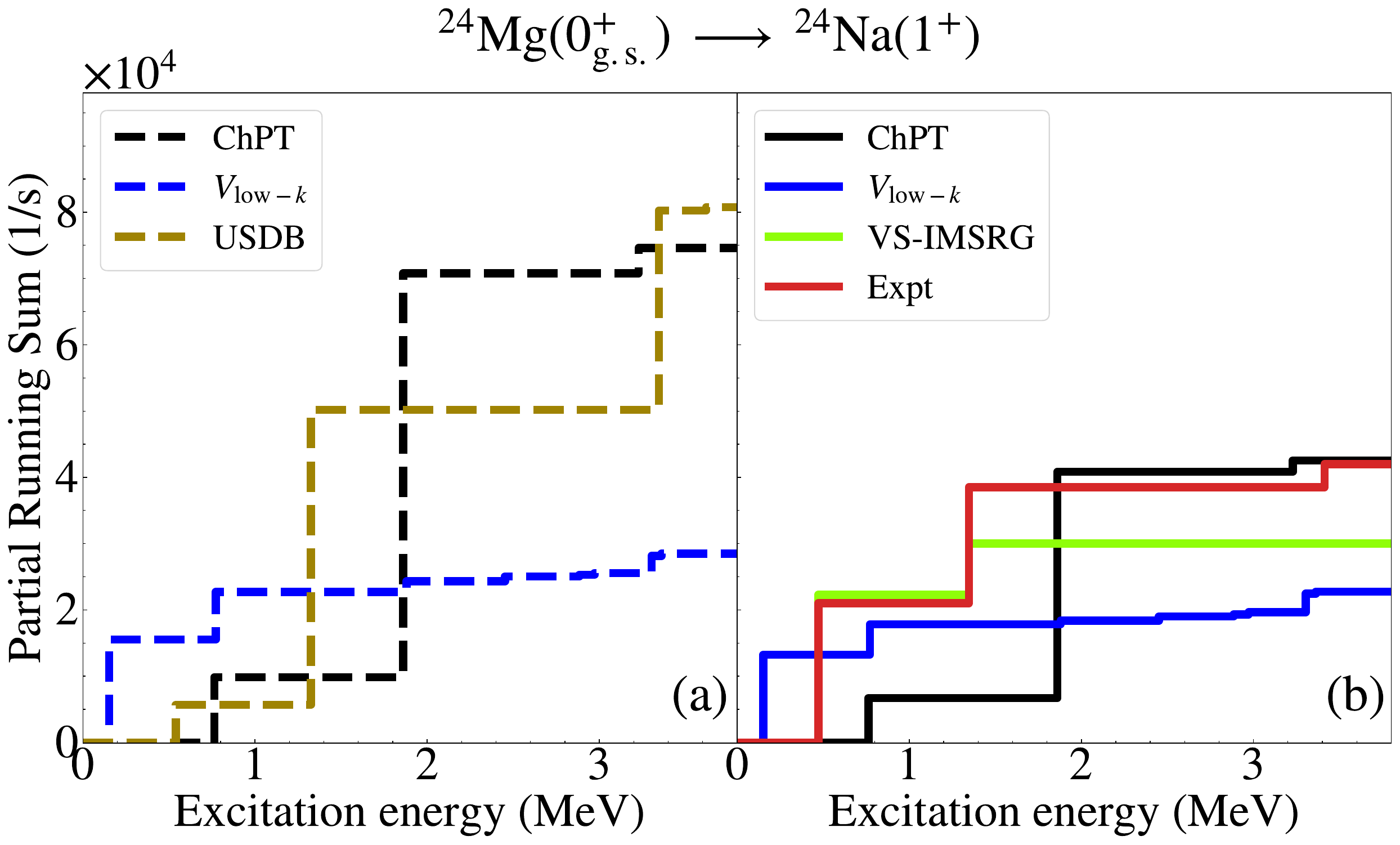}
    \caption{Same as in Fig. \ref{fig:rsum_Na_12}, but for the capture rate from the ground state of ${}^{24}$Mg to the ${}^{24}$Na$(1^+)$ final states. In panel (b) they are reported -- in green -- also the results obtained with VS-IMSRG \cite{Jokiniemi21}}
    \label{fig:rsum_Mg_1}
\end{figure}

\begin{figure}[h!]
    \centering
    \includegraphics[scale=0.2]{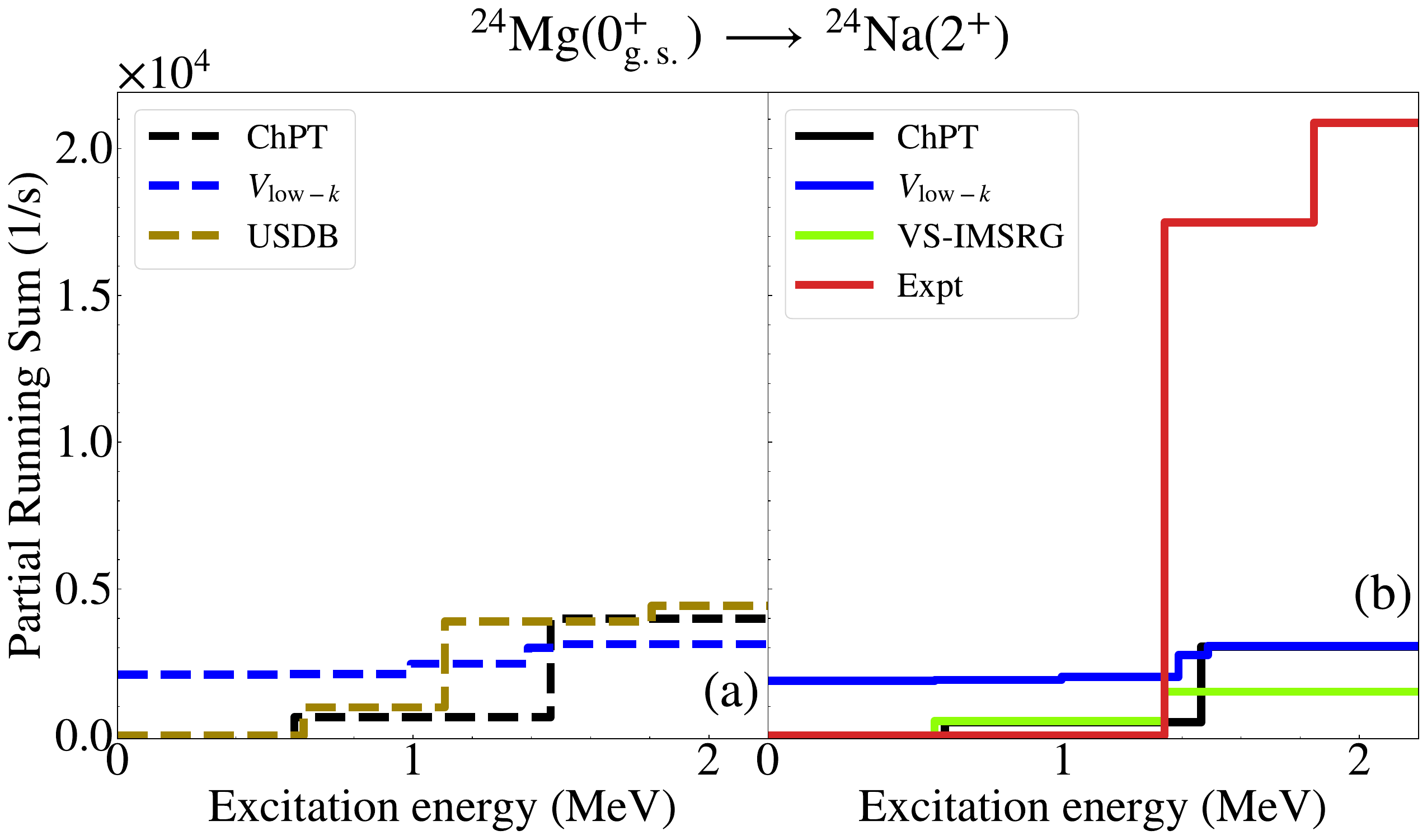}
    \caption{Same as in Fig. \ref{fig:rsum_Mg_1}, but for the capture rate from the ground state of ${}^{24}$Mg to the ${}^{24}$Na$(2^+)$ final states.}
    \label{fig:rsum_Mg_2}
\end{figure}

\begin{figure}[h!]
    \centering
    \includegraphics[scale=0.2]{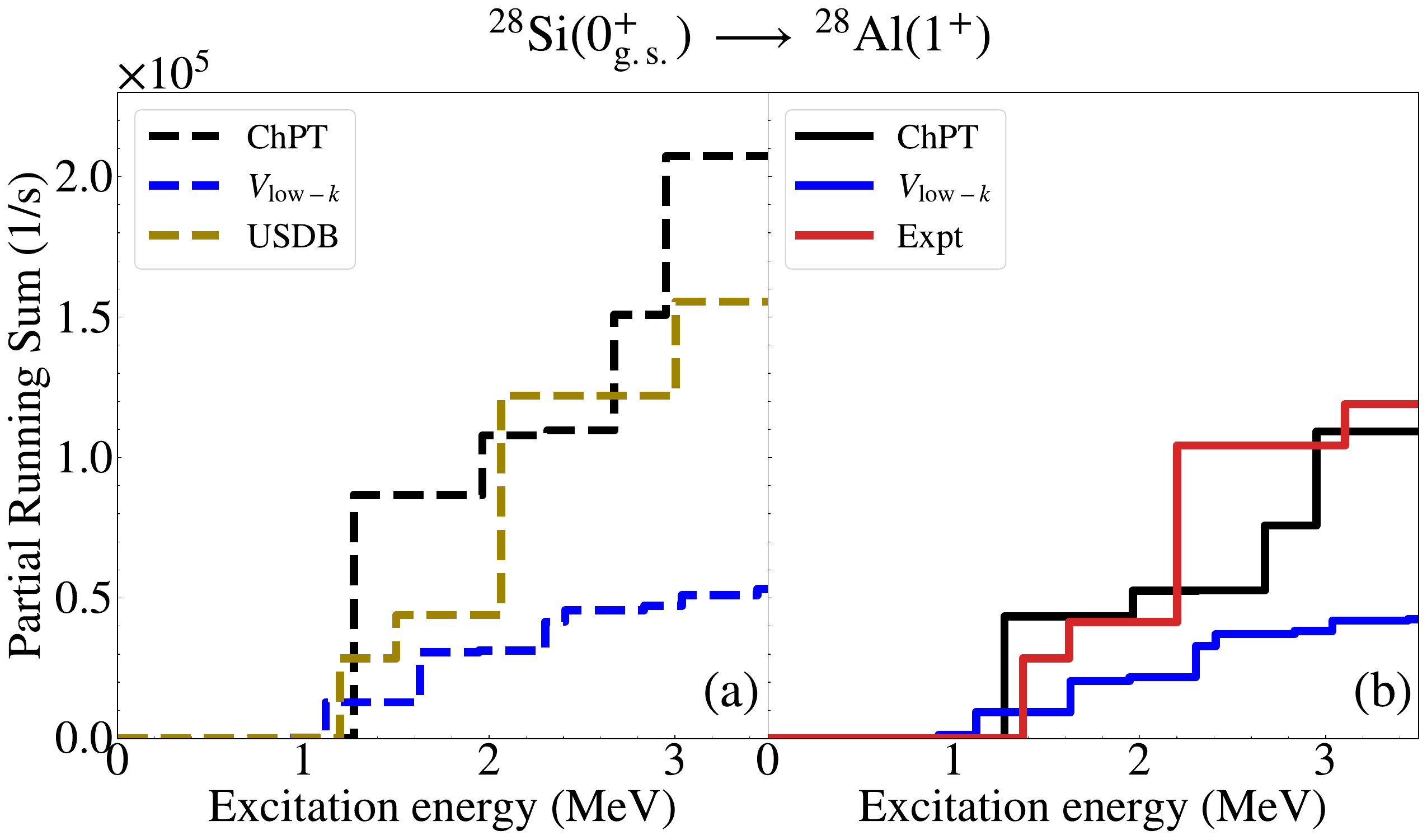}
    \caption{Same as in Fig. \ref{fig:rsum_Na_12}, but for the capture rate from the ground state of ${}^{28}$Si to the ${}^{28}$Al$(1^+)$ final states.}
    \label{fig:rsum_Si_1}
\end{figure}

\begin{figure}[h!]
    \centering
    \includegraphics[scale=0.2]{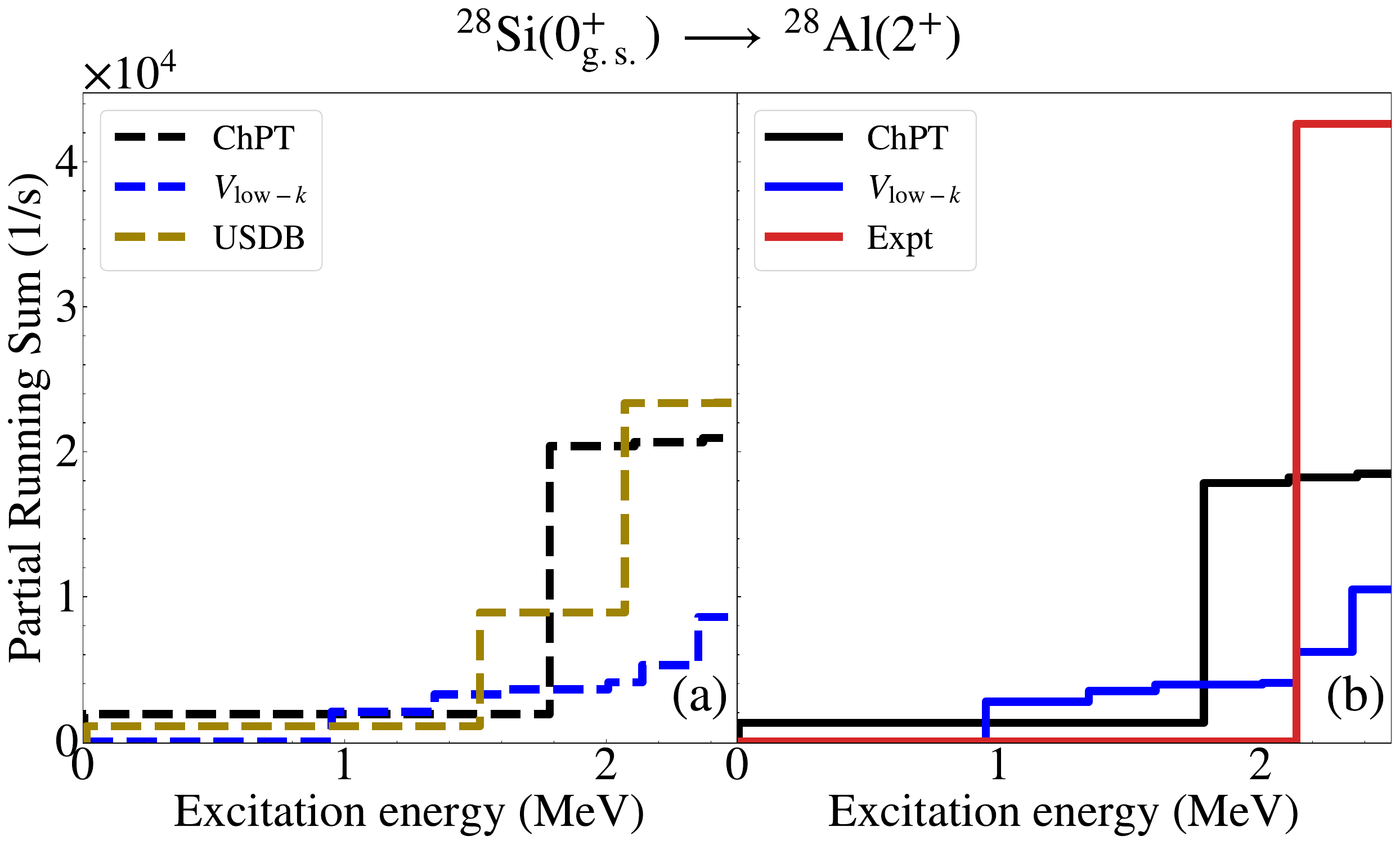}
    \caption{Same as in Fig. \ref{fig:rsum_Na_12}, but for the capture rate from the ground state of ${}^{28}$Si to the ${}^{28}$Al$(2^+)$ final states.}
    \label{fig:rsum_Si_2}
\end{figure}

\begin{figure}[h!]
    \centering
    \includegraphics[scale=0.2]{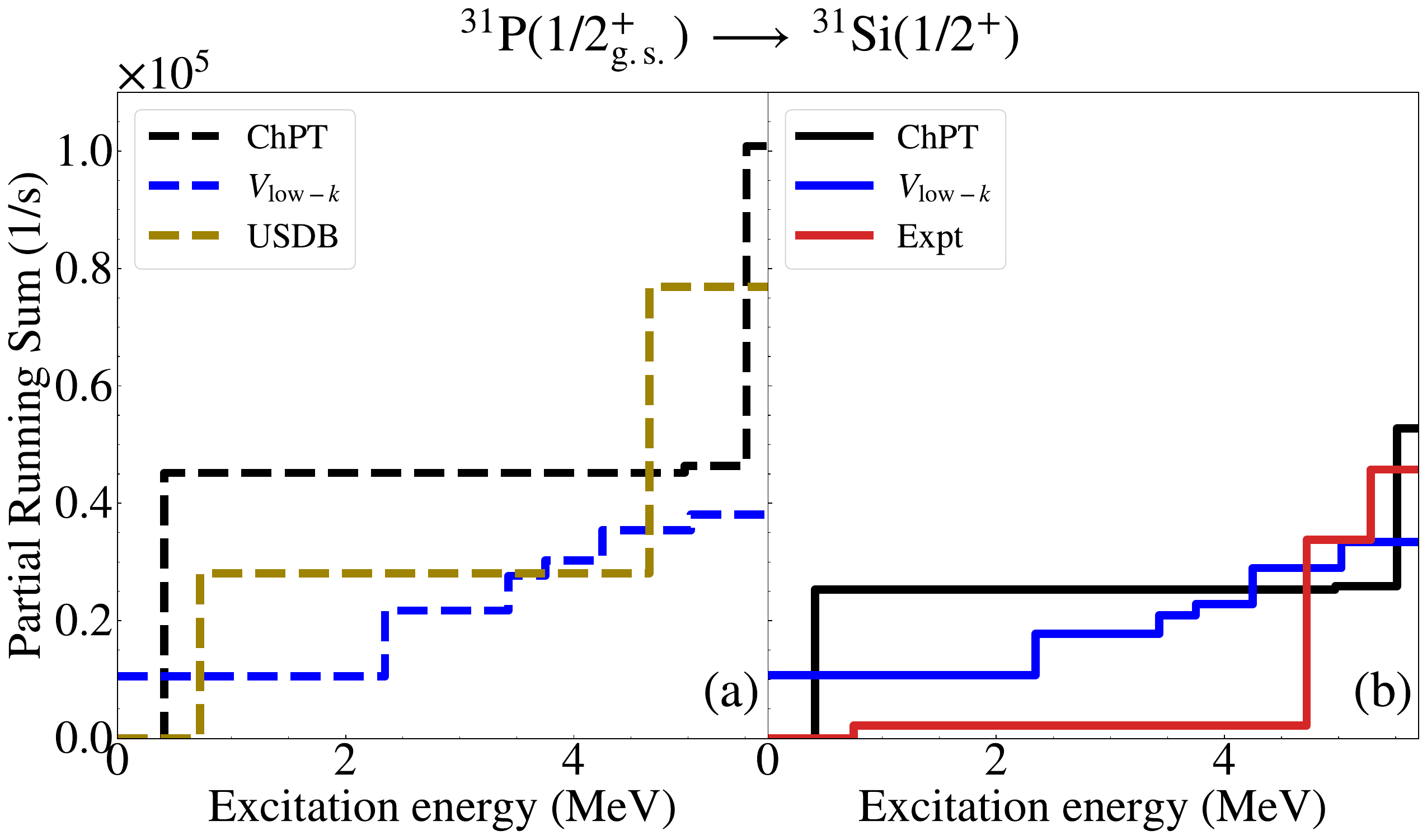}
    \caption{Same as in Fig. \ref{fig:rsum_Na_12}, but for the capture rate from the ground state of ${}^{31}$P to the ${}^{31}$Si$(1/2^+)$ final states.}
    \label{fig:rsum_P_12}
\end{figure}

\begin{figure}[h!]
    \centering
    \includegraphics[scale=0.2]{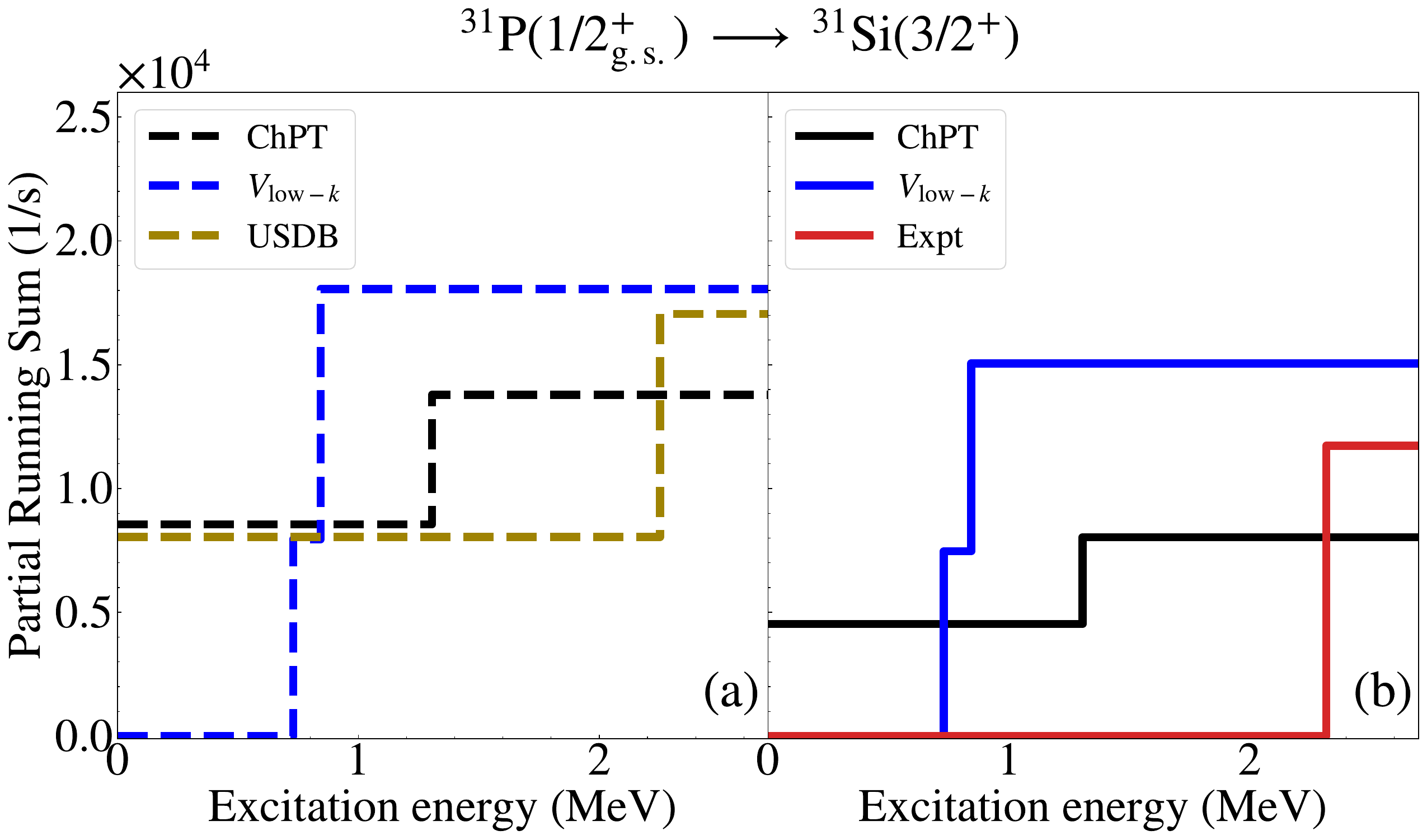}
    \caption{Same as in Fig. \ref{fig:rsum_Na_12}, but for the capture rate from the ground state of ${}^{31}$P to the ${}^{31}$Si$(3/2^+)$ final states.}
    \label{fig:rsum_P_32}
\end{figure}

\begin{figure}[h!]
    \centering
    \includegraphics[scale=0.2]{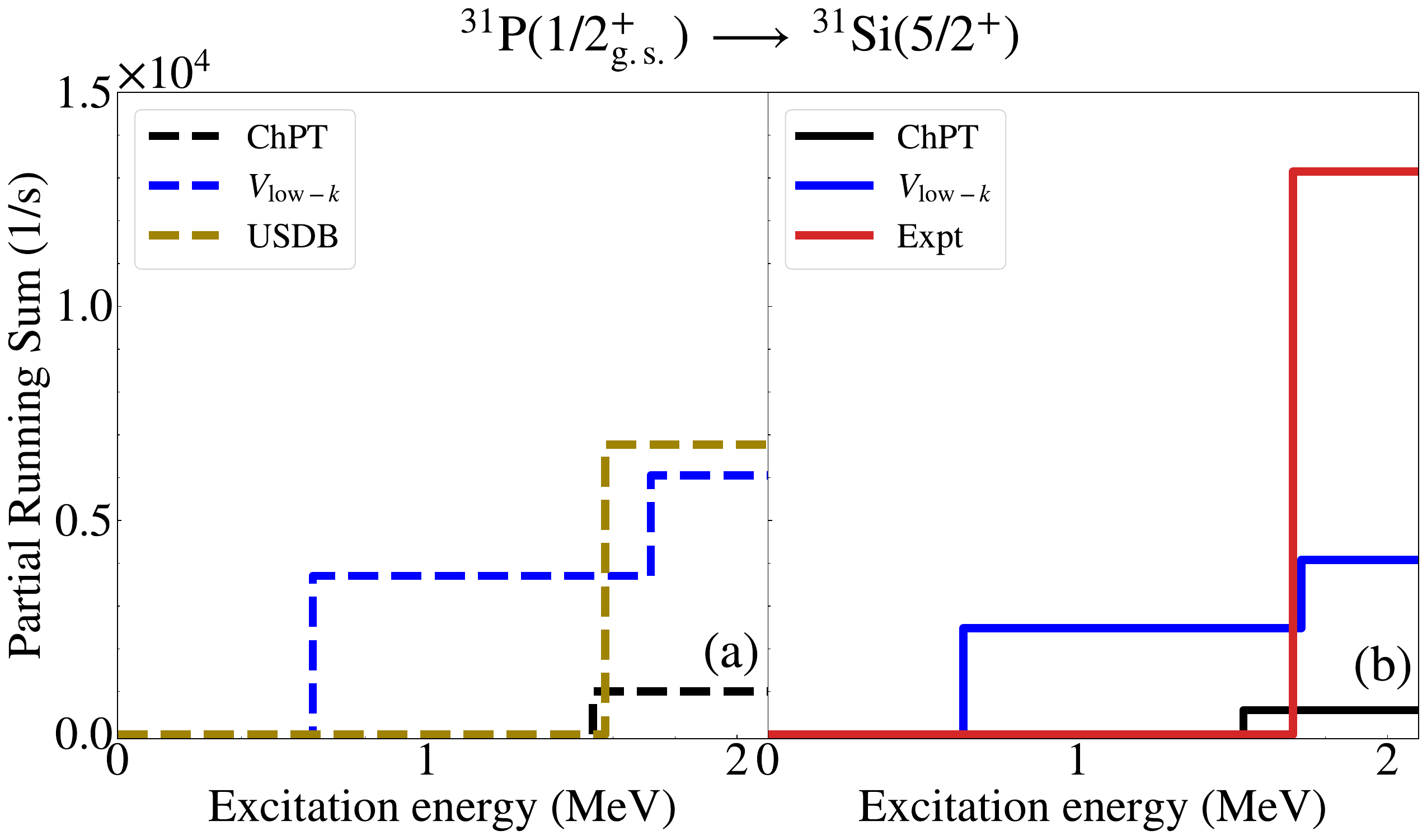}
    \caption{Same as in Fig. \ref{fig:rsum_Na_12}, but for the capture rate from the ground state of ${}^{31}$P to the ${}^{31}$Si$(5/2^+)$ final states.}
    \label{fig:rsum_P_52}
\end{figure}

\begin{figure}[h!]
    \centering
    \includegraphics[scale=0.2]{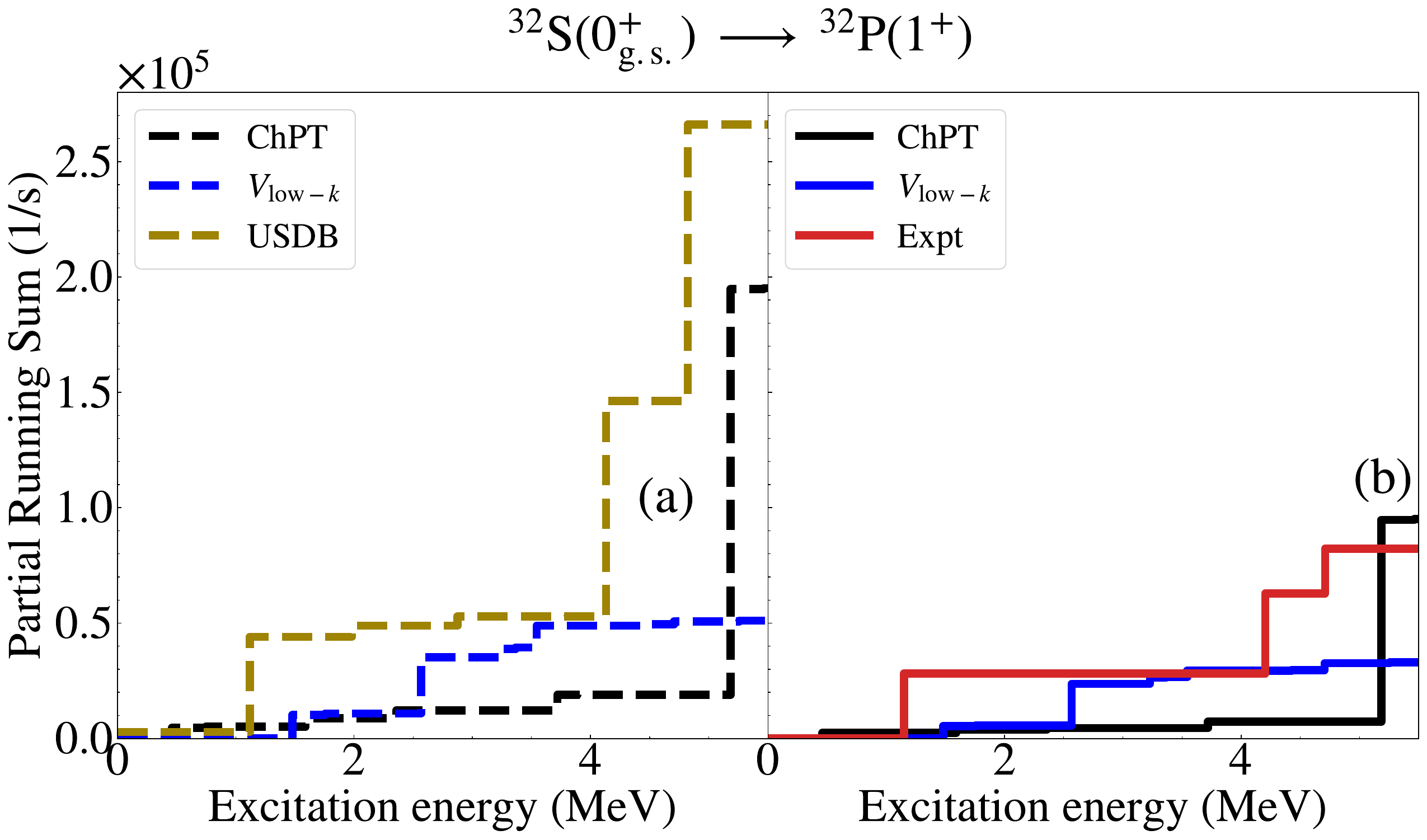}
    \caption{Same as in Fig. \ref{fig:rsum_Na_12}, but for the capture rate from the ground state of ${}^{32}$S to the ${}^{32}$P$(1^+)$ final states.}
    \label{fig:rsum_S_1}
\end{figure}

\begin{figure}[h!]
    \centering
    \includegraphics[scale=0.2]{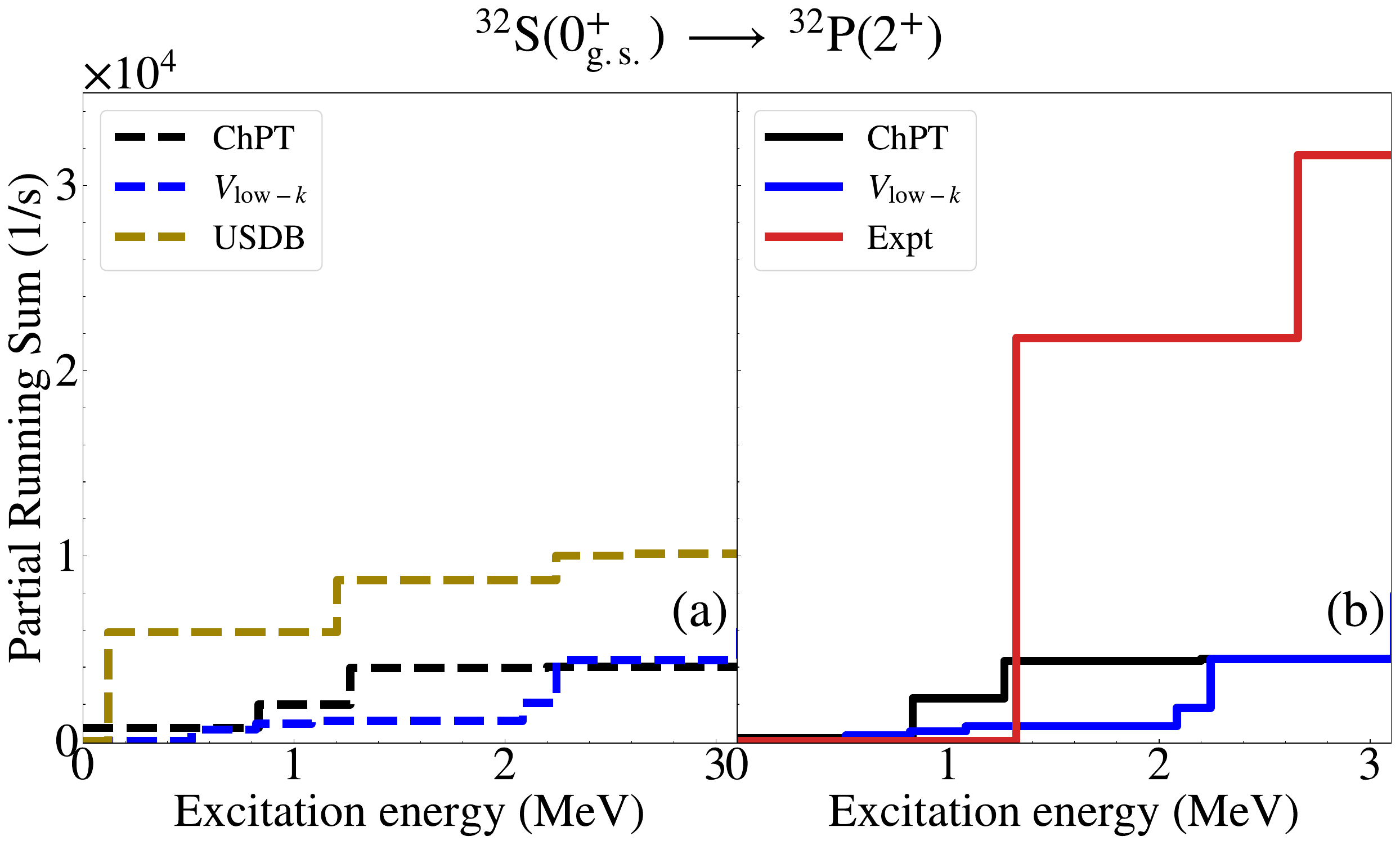}
    \caption{Same as in Fig. \ref{fig:rsum_Na_12}, but for the capture rate from the ground state of ${}^{32}$S to the ${}^{32}$P$(2^+)$ final states.}
    \label{fig:rsum_S_2}
\end{figure}

We now focus on the partial muon capture rates of the nuclei considered in this work, namely ${}^{23}$Na (Figs.~\ref{fig:rsum_Na_12},~\ref{fig:rsum_Na_32},~\ref{fig:rsum_Na_52}), ${}^{24}$Mg (Figs.~\ref{fig:rsum_Mg_1},~\ref{fig:rsum_Mg_2}), ${}^{28}$Si (Figs.~\ref{fig:rsum_Si_1},~\ref{fig:rsum_Si_2}), ${}^{31}$P (Figs.~\ref{fig:rsum_P_12},~\ref{fig:rsum_P_32},~\ref{fig:rsum_P_52}) and ${}^{32}$S (Figs.~\ref{fig:rsum_S_1},~\ref{fig:rsum_S_2}).

We point out that, in our calculations, as for example in \cite{Siiskonen:1998vdi}, for the two odd-mass nuclei, ${}^{23}$Na and ${}^{31}$P, the hyperfine structure of the muonic atom is not taken into account since we start from Morita-Fujii formulation \cite{Morita60} of the OMC.
In this formulation, in fact, a statistical population of the hyperfine states is assumed.

In all the figures, we start by comparing in panel (a) the partial capture rates obtained with the bare operator using the chiral and $V_\mathrm{low\!-\!k}$ interactions with those calculated starting from the phenomenological USDB effective interaction~\cite{Brown06b} (Figs. \ref{fig:rsum_Na_12}-\ref{fig:rsum_S_2}).

This comparison reveals a remarkable agreement between the results obtained with the chiral EFT potential and those from USDB. In contrast, in most cases, the partial rates obtained with \vlwk~ are systematically smaller. 
This behavior traces back to the connection between spectroscopic properties and capture rates, indeed, the chiral potential yields an agreement with experiment that is comparable to one that is provided by the USDB interaction, as regards the low-lying excitation spectra and transition strengths.

In panel (b) of Figs. \ref{fig:rsum_Na_12}-\ref{fig:rsum_S_2}, we present the partial capture rates calculated using the effective decay operators for the chiral and \vlwk~ interactions, and compare them with the available data \cite{Johnson96,Gorringe99}.
Additionally, for $^{24}$Mg, we report the results of Ref.~\cite{Jokiniemi21}, obtained within the framework of the {\it ab initio} VS-IMSRG approach.
It should be pointed out that we refer to VS-IMSRG results which are obtained without the contribution of two-body electroweak currents.

Overall, the results obtained using the chiral \heff~ provide a better description of the experiment with respect to those scored with the \vlwk, mirroring our observation about the reproduction of the low-energy spectroscopic properties.
Moreover, we have found that the calculated muon-capture rates exhibit little sensitivity to the values of the momentum transfer.
Therefore, the differences among the results with different \heffs~ can be mainly ascribed to the many-body wave functions, rather than to differences in eigenvalues.

A less satisfactory description occurs for the capture from the ground state of ${}^{23}$Na to the $1/2^+$ states of $^{23}$Ne, and the transitions of $^{24}$Mg,$^{32}$S to the $2^+$ states of $^{24}$Na,$^{32}$P, respectively. 
The experimental OMC strengths are underestimated by the calculations with our \heffs~ as well as by the USDB phenomenological interaction, indicating that these discrepancies are not due to choice of the effective Hamiltonian.
It is worth noting that our results for the OMC of $^{24}$Mg (see panel (b) in Fig. 14) are consistent with those reported in Ref. \cite{Jokiniemi21}, where calculations have been carried out by way of the {\it ab initio} VS-IMSRG approach.
This agreement between RSM and {\it ab initio} results -- that occurs both for $1,2^+$ states of $^{24}$Na -- testifies that the discrepancy between theory and experiment is not related to missing many-body configurations in our approach, and needs further investigations.

\section{Summary and outlook}
\label{conclusions}
In this work we have spotted the focus on the study of ordinary muon capture within the framework of realistic nuclear shell model, namely by constructing effective Hamiltonians and decay operators through the many-body perturbation theory, and starting from realistic nuclear Hamiltonians.
The interest of such an investigation stems from the fact that OMC induces a high‑momentum ($\simeq$ 100 MeV/c) charged‑current response in the nucleus
that is comparable to the momentum transfer encountered in \zbb~ decay, and therefore represents a challenging testing ground to assess nuclear models that aim to predict reliable nuclear matrix elements for such an unobserved nuclear decay.
The nuclear systems we have considered for our analysis are a large set of nuclei belonging to the $sd$ shell, that provide a relevant amount of observed OMC rates to compare with our theoretical calculations.

We have employed two different effective SM Hamiltonians, one constructed from a low-momentum nucleon-nucleon interaction derived from the high-precision CD-Bonn potential, and another obtained from a $2N$ plus $3N$ nuclear Hamiltonian rooted in the chiral perturbation theory. 
This has allowed us to highlight a strong correlation between the ability of the effective Hamiltonian to reproduce spectroscopic properties and its performance in OMC observables.
Namely, one of our main outcomes is that the better the calculated spectroscopic observables, the closer the agreement with experimental capture rates.

We have also been able to benchmark our approach to the OMC with an {\it ab initio} method, the so-called VS-IMSRG approach \cite{Jokiniemi21}, and the agreement between the results of the two calculations testifies the ability of effective Hamiltonians and decay operators derived from perturbative many-body theory to recover the contributions of nuclear configurations which are not explicitly included in a truncated shell-model space.

The overall agreement with experiment, especially with ChPT \heff, is satisfactory, and the analysis we have reported indicates that nuclei, whose low‑energy spectroscopic properties are well reproduced by this \heff, also yield reliable predictions for high‑momentum weak processes.

As future perspectives of the development of the study of OMC within the framework of the realistic shell model there are two major points that are worth to underline.

The first one is the theoretical improvements that should be investigated, and we identify three possible items that it is worth to consider: one is to evaluate the impact of two-body electroweak currents – accounting at subleading order for the quark structure of the nucleons –, that can be derived consistently with the nuclear Hamiltonian in the framework of the chiral perturbation theory (see for example Ref. \cite{Marcucci18}).
  We have already investigated recently the impact of two-body electroweak currents on the calculation of Gamow-Teller observables, and we have observed that these contributions could lead to a significant improvement of the reproduction of the data \cite{Coraggio24}.
  Another action that for sure would improve the quality of the nuclear wave functions is the enlargement of the model space, namely including also the degrees of freedom of the orbitals of the $0f1p$ shell.
This would also lead to a lesser impact of the renormalization of the decay operators, but it is very challenging from the point of view of the derivation of the SM effective interaction within model spaces that are characterized by two major shells, since this leads to the introduction of poles in the denominators of the \qbox~ perturbative expansion \cite{Suzuki11}.
Moreover, a systematic study of the effect of statistical population of hyperfine states of the muonic atom  will be performed in the near future to assess the reliability of this approximation.

The second one is our hope that in a near future more experimental efforts  about OMC could lead to a larger amount of data, especially in mass regions where candidates to the observation of \zbb~ decay belong.
In particular high‑precision OMC experiments, such as MONUMENT \cite{Araujo24}, will also be critical to further constrain nuclear interactions and operator renormalization schemes.
This would be an important achievement also for the theoretical community, since it would provide more helpful data to test nuclear theory and models towards the calculation of reliable \zbb~ nuclear matrix elements. 

\section{Acknowledgments}
This work was supported in part by the Italian Ministero
dell’Universit\`a e della Ricerca (MUR), through the grant Progetto di Ricerca d'Interesse Nazionale 2022 (PRIN 2022) ``Exploiting Separation of Scales in Nuclear Structure and Dynamics'', CUP B53D23005070006, and by the JSPS KAKENHI Grant Numbers JP21K13919 and JP23KK0250.

\bibliographystyle{apsrev}
\bibliography{biblio}

\end{document}